\documentclass[fleqn,usenatbib]{mnras}

\usepackage[T1]{fontenc}

\DeclareRobustCommand{\VAN}[3]{#2}
\let\VANthebibliography\thebibliography
\def\thebibliography{\DeclareRobustCommand{\VAN}[3]{##3}\VANthebibliography}

\usepackage{graphicx}	
\usepackage{amsmath}	
\usepackage{amssymb}	
\usepackage{newtxtext,newtxmath}
\usepackage{multirow}
\usepackage{lscape}
\usepackage{makecell}
\usepackage{subfigure}
\usepackage{float}
\usepackage{newtxtext,newtxmath}

\makeatletter
\newcommand{\labtext}[2]{%
  \@bsphack
  \csname phantomsection\endcsname 
  \def\@currentlabel{#1}{\label{#2}}%
  \@esphack
}
\makeatother

\title[MWL periodicity search in $\gamma$-ray blazars]{Multiwavelength periodicity search in a sample of $\gamma$-ray bright blazars}

\author[J. Otero-Santos et al.]{
J. Otero-Santos,$^{1,2}$\thanks{E-mail: joteros@iac.es (JOS)}
P. Peñil,$^{3}$
J. A. Acosta-Pulido,$^{1,2}$
J. Becerra González,$^{1,2}$
C. M. Raiteri,$^{4}$
\newauthor
M. I. Carnerero$^{4}$ and
M. Villata$^{4}$
\\
$^{1}$Instituto de Astrof\'isica de Canarias (IAC), E-38200 La Laguna, Tenerife, Spain\\
$^{2}$Universidad de La Laguna (ULL), Departamento de Astrof\'isica, E-38206 La Laguna, Tenerife, Spain\\
$^{3}$Department of Physics and Astronomy, Clemson University, Kinard Lab of Physics, Clemson, SC 29634-0978, USA\\
$^{4}$INAF-Osservatorio Astrofisico di Torino, via Osservatorio 20, 10025 Pino Torinese, Italy\\
}

\date{Accepted XXX. Received YYY; in original form ZZZ}

\pubyear{2015}

\begin{document}
\label{firstpage}
\pagerange{\pageref{firstpage}--\pageref{lastpage}}
\maketitle

\begin{abstract}
We present the results of a long-term periodicity search in a sample of $\gamma$-ray blazars within a multiwavelength context. These blazars have been selected from the Steward Observatory sample as part of its optical monitoring program between 2008 and 2018. We study 15 sources with a temporal coverage in their optical total and polarized emission sufficiently large ($>9$~years) to perform a reliable long-term periodicity analysis. We collect data from several observatories to extend the coverage, enabling the search of longer periods. In addition, data are also gathered in the high-energy ($E>100$~MeV) $\gamma$-ray band from the \textit{Fermi} Large Area Telescope; and in the 15-GHz radio band from the Owens Valley Radio Observatory. We identify 5 promising candidates to host {quasi-periodic} emission, AO~0235+164, PKS~1222+216, Mrk~501, BL~Lacertae and 1ES~2344+514 with periods in one or more bands and statistical significances $\sim$3$\sigma$ after trial factor correction. AO~0235+164 shows a period of $\sim$8.2~years in the R band; PKS~1222+216 has a {quasi-periodic} modulation in its total and polarized optical emission of $\sim$1.6~years; Mrk~501 displays a $\sim$5-year {quasi-periodicity} in optical and radio wavelengths; BL~Lacertae presents a period of $\sim$1.8~years in its polarized emission; and 1ES~2344+514 shows a hint of a $\sim$5.5-year period in its optical R band. We interpret these results in the framework of the most common models and scenarios, namely the presence of a binary supermassive black hole system; or geometrical effects like helical or precessing jets.

\end{abstract}




\section{Introduction}
Blazars, active galactic nuclei (AGNs) with a relativistically boosted jet closely aligned with the line of sight of the observer, are among the most violent and variable objects in the Universe. They can emit in every wavelength of the electromagnetic spectrum, from radio to very-high-energy (VHE, E>100 GeV) $\gamma$ rays. Blazars are typically subdivided in two classes according to their optical spectrum \citep{urry1995}. The optical spectrum from BL Lac objects is dominated by continuum emission with almost no spectral features. On the other hand, flat spectrum radio quasars (FSRQs) display bright and broad emission lines, with equivalent widths $|\text{EW}|>5$ \AA \ in the rest frame \citep[see e.g.][]{giommi2012}.

These violent sources display very strong variability in many different time scales: as short as a few minutes \citep[intraday variability, see e.g.][]{aleksic2011,bhatta2018,raiteri2021a,raiteri2021b}, variability in time scales of a few days or weeks \citep[short-term variability, for instance][]{gupta2008,rani2010,sandrinelli2014} and changes on the order of several months or years \citep[long-term variability, see for example][]{carnerero2017,nalewajko2019,sarkar2019}. Although this variability is commonly attributed to stochastic processes and unpredictable flare events, there have been several claims in the last years of detection of quasi-periodic oscillations (QPOs) with high statistical significance. These possible QPO candidates have been observed in all wavelengths, for instance \cite{tripathi2021} in radio, \cite{sandrinelli2016} in optical, \cite{wang2017} in X rays or \cite{penil2020} in $\gamma$ rays; and in different time scales, like e.g. \cite{lachowicz2009,rani2010b} in time scales of minutes or hours, \cite{zhou2018} in time scales of several days, \cite{kushwaha2020} in scales of months or \cite{penil2020} in time scales of a few years.

Variability is one of the main tools to study and understand the physics involved in one of the most energetic objects of the Universe. Variability studies can help to interpret the relativistic emission of blazars in the framework of different scenarios depending on its nature and variability time scale. In particular, periodic variability has been interpreted in the framework of several models that could help to understand the behaviour of blazars. One of the most extended interpretations is that QPOs may be driven by a binary system of supermassive black holes (SMBH). The two most promising candidates for hosting a binary SMBH system are OJ~287 \citep{sillanpaa1988} and PG~1553+113 \citep{tavani2018}, and several candidates have been proposed in recent studies \citep[for instance 3C 454.3 by][]{qian2021}. However, the true nature of these objects is still under debate, and often interpreted under scenarios not involving a binary system \citep[e.g. OJ~287 by][]{butuzova2020}.


Alternatively, QPOs have also been interpreted in the framework of different geometrical models such as the precession of the jet \citep[lighthouse effect,][]{camenzind1992}, helical jets or helical trajectories along the jet \citep{rieger2004}, twisted and inhomogeneous jets \citep{raiteri2017} or orbital movement of blob in the jet under the influence of the magnetic field \citep{mohan2015}. Furthermore, periodic variations can also be caused by instabilities introduced by the accretion disc in the jet \citep[e.g.][]{dong2020}. Thus, periodicity studies are a powerful tool for learning about the emitting mechanisms in blazars and the origin of their variability. 

In this paper we present the multiwavelength (MWL) long-term periodicity search performed on 15 $\gamma$-ray bright blazars monitored for $\sim$10 years by the Steward Observatory in Arizona, in support of the \textit{Fermi} $\gamma$-ray survey. The paper is structured as follows: in Section \ref{sec2} the optical and MWL data used in this work are introduced; in Section \ref{sec3} we describe the analysis tools employed; in Sections \ref{sec4} and \ref{sec5} the results of the analysis and the discussion are presented; and in Section \ref{sec6} we summarise the main conclusions of the work.

\section{Multiwavelength data}\label{sec2}
We have analysed a sample of 15 $\gamma$-ray bright blazars monitored by the Steward Observatory for approximately 10 years, between 2008 and 2018\footnote{\url{http://james.as.arizona.edu/~psmith/Fermi/}}. The candidates were selected according to the temporal coverage in both the total and polarized light, among all the blazars observed by this program. We have selected those objects with a time coverage larger than 9 years to permit the performance of a reliable long-term periodicity analysis. We collected MWL data from different observatories around the globe in radio, optical and $\gamma$-ray frequencies to study the periodicity in different bands. The details of each data sample are discussed in the following sections. An example of the MWL LCs is presented in Figures \ref{ao0235_lcs}, \ref{pks1222_lcs}, \ref{mrk501_lcs}, \ref{bllac_lcs} and \ref{1es2344_lcs}. The rest of the data are included as online material (Figures \ref{oj248lc} to \ref{3c454lc}).

\subsection{Optical data}
\subsubsection{Photometry}
We have performed the long-term periodicity analysis on the R- and V- band light curves (LCs) of the blazars included in this paper. The data sample is based on the photometric data provided by the Steward Observatory between 2008 and 2018 \citep{smith2009}. Then, in order to obtain LCs with enough temporal coverage, we collected data from several ground-based optical facilities and observatories. The Whole Earth Blazar Telescope\footnote{\url{https://www.oato.inaf.it/blazars/webt/}} \citep[WEBT,][]{villata2009} Collaboration provided R- and V- band data for several targets. Furthermore, we make use of the data in the R and V wave bands taken by the Small and Moderate Aperture Research Telescope System\footnote{\url{http://www.astro.yale.edu/smarts/glast/home.php}} \citep[SMARTS,][]{bonning2012} consortium, that observes several blazars monitored by the \textit{Fermi} Large Area Telescope (\textit{Fermi}-LAT). We also made use of the publicly available R-band data from the Katzman Automatic Imaging Telescope\footnote{\url{http://herculesii.astro.berkeley.edu/kait/agn/}} \citep[KAIT,][]{li2003}, that has been monitoring 163 bright \textit{Fermi}-LAT blazars since 2011. Additionally, the data obtained from the American Association of Variable Star Observers (AAVSO) International Database\footnote{\url{https://www.aavso.org}} in both R and V filters. Public R-band data from the Tuorla blazar monitoring\footnote{\url{https://users.utu.fi/kani/1m/}} available at the VizieR Online Data Catalog \citep{nilsson2018}. R-band data from the IAC80 telescope located at the Observatorio del Teide, the STELLar Activity Robotic Observatory\footnote{\url{https://www.aip.de/en/research/projects/stella/}} \citep[STELLA,][]{strassmeier2004} and the Las Cumbres Observatory\footnote{\url{https://lco.global}} \citep[LCOGT,][]{brown2013} were included in the analysis. Finally, V-band data from the Catalina Real-Time Transient Survey\footnote{\url{http://crts.caltech.edu/index.html}} \citep[CRTS,][]{drake2009} and the All-Sky Automated Survey for Supernovae\footnote{\url{https://asas-sn.osu.edu}} \citep[ASAS-SN,][]{shappee2014,kochanek2017} were also used in this work. After collecting all the optical data, we end up with 15 R- and V-band optical LCs with typical time coverage $\gtrsim$12 years for most of the blazars included in this work. The sources analysed here and the time span of their LCs are listed in Table \ref{table_observations}. For those sources displaying a bright host galaxy emission (i.e. Mrk~421, Mrk~501 and 1ES~2344+514), we subtracted the stellar population contribution following the procedure from \cite{nilsson2007}.

\subsubsection{Polarimetry}
To complement the collected photometric data, we have made use of the polarimetric observations performed by the Steward Observatory. These data were obtained simultaneously to the photometric observations. The polarized magnitude was obtained from the total magnitude using the polarization degree measured in the band of 5000-7000 \AA. The polarization degree was obtained from the Stokes parameters as $P=100 \cdot \sqrt{<q>^{2}+<u>^{2}}$. $P$ has been corrected from the statistical bias associated to the fact that the polarization degree must be a positive physical quantity, following the prescription from \cite{wardle1974}. The polarimetric data available for the selected blazars cover an interval of almost 10 years for all the blazars of the sample (see Table \ref{table_observations}). 

\begin{table*}
\centering
\caption{List of blazars included in the periodicity analysis based on their optical R- and V-band and optical polarization data available. All the LCs were daily binned except from the \textit{Fermi}-LAT LCs retrieved from the LCR that correspond to a 30-day binning.}
\label{table_observations}
\resizebox{\textwidth}{!}{%
\begin{tabular}{cccccccccccc}
\hline
\multirow{3}{*}{Source} & \multirow{3}{*}{Type} & \multicolumn{2}{c}{Optical R band} & \multicolumn{2}{c}{Optical V band} & \multicolumn{2}{c}{Optical polarization} & \multicolumn{2}{c}{HE $\gamma$-ray band} & \multicolumn{2}{c}{15-GHz radio band} \\ \cline{3-12} 
 &  & \multirow{2}{*}{N$_{\text{obs}}$} & Date & \multirow{2}{*}{N$_{\text{obs}}$} & Date & \multirow{2}{*}{N$_{\text{obs}}$} & Date & \multirow{2}{*}{N$_{\text{obs}}$} & Date & \multirow{2}{*}{N$_{\text{obs}}$} & Date \\
 &  &  & [MJD] &  & [MJD] &  & [MJD] &  & [MJD] &  & [MJD] \\ \hline
AO 0235+164 & BL Lac & 10462 & 42689 -- 58869 & 1046 & 53620 -- 58161 & 176 & 54743 -- 58161 & 165 & 54687 -- 59617 & 692 & 54471 -- 59025 \\ \hline
OJ 248 & FSRQ & 662 & 53797 -- 58862 & 831 & 53470 -- 58254 & 224 & 54743 -- 58254 & 165 & 54687 -- 59617 & 532 & 54628 -- 59022 \\ \hline
OJ 287 & BL Lac & 4198 & 52615 -- 59475 & 8516 & 51879 -- 59476 & 510 & 54743 -- 58292 & 165 & 54687 -- 59617 & 636 & 54473 -- 59022 \\ \hline
Mrk 421 & BL Lac & 9748 & 52330 -- 59167 & 6198 & 50217 -- 59351 & 562 & 54744 -- 58306 & 165 & 54687 -- 59617 & 840 & 54473 -- 59030 \\ \hline
W Comae & BL Lac & 536 & 54622 -- 58306 & 2043 & 53396 -- 59388 & 334 & 54768 -- 58306 & 165 & 54687 -- 59617 & 697 & 54473 -- 59028 \\ \hline
PKS 1222+216 & FSRQ & 784 & 54948 -- 58844 & 1045 & 53469 -- 58449 & 418 & 54948 -- 58306 & 165 & 54687 -- 59617 & -- & -- \\ \hline
3C 273 & FSRQ & 1120 & 54491 -- 58989 & 10017 & 53464 -- 59390 & 302 & 54795 -- 58305 & 165 & 54687 -- 59617 & 786 & 54473 -- 59021 \\ \hline
3C 279 & FSRQ & 4049 & 53083 -- 58963 & 4078 & 51992 -- 59394 & 500 & 54795 -- 58306 & 165 & 54687 -- 59617 & 607 & 54473 -- 59021 \\ \hline
PKS 1510-089 & FSRQ & 2426 & 54272 -- 59001 & 1745 & 53493 -- 59444 & 382 & 54830 -- 58306 & 165 & 54687 -- 59617 & -- & -- \\ \hline
Mrk 501 & BL Lac & 3730 & 52529 -- 59278 & 1111 & 53526 -- 58380 & 475 & 54745 -- 58306 & 165 & 54687 -- 59617 & 615 & 54854 -- 59027 \\ \hline
1ES 1959+650 & BL Lac & 1127 & 52528 -- 58409 & 1421 & 54745 -- 59175 & 145 & 54745 -- 58293 & 165 & 54687 -- 59617 & 766 & 54475 -- 59027 \\ \hline
BL Lacertae & BL Lac & 30045 & 42753 -- 59460 & 17621 & 39974 -- 59475 & 742 & 54743 -- 58312 & 165 & 54687 -- 59617 & -- & -- \\ \hline
CTA 102 & FSRQ & 11799 & 52433 -- 58425 & 4958 & 53520 -- 59454 & 292 & 55124 -- 58306 & 165 & 54687 -- 59617 & -- & -- \\ \hline
3C 454.3 & FSRQ & 16558 & 48103 -- 59171 & 4985 & 45229 -- 59461 & 569 & 54743 -- 58306 & 165 & 54687 -- 59617 & 1112 & 54471 -- 59028 \\ \hline
1ES 2344+514 & BL Lac & 995 & 52528 -- 59018 & 422 & 54743 -- 58447 & 198 & 54743 -- 58282 & 165 & 54687 -- 59617 & -- & -- \\ \hline

\end{tabular}%
}
\end{table*}

\subsection{High-energy $\gamma$ rays: \textit{Fermi}-LAT data}
The pair-conversion Large Area Telescope (LAT) on board the \textit{Fermi} satellite observes the entire sky every three hours thanks to its survey mode. It operates in the high-energy (HE) $\gamma$-ray band, in an energy range from 50 MeV up to >300 GeV \citep{atwood2009}. We have made use of the LCs available at the \textit{Fermi}-LAT public Light Curve Repository (LCR)\footnote{\url{https://fermi.gsfc.nasa.gov/ssc/data/access/lat/LightCurveRepository/about.html}}. The LCR is a LC database for all the variable sources included in the 4FGL-DR2 catalog according to their variability index \citep{abdollahi2020}. It performs the LC analysis with the standard \textit{Fermi}-LAT analysis tools, with the instrument response function \texttt{P8R3\_SOURCE\_V3}. It uses a maximum likelihood analysis \citep[see e.g.][]{abdo2009}, considering a region of interest (ROI) of a 12$^{\circ}$ radius and an energy range between 100~MeV and 100~GeV. A maximum zenith angle cut of 90$^{\circ}$ is also applied to the data to avoid introducing contamination from the limb of the Earth. The model used for the likelihood analysis includes all the sources within a radius of 30$^{\circ}$ centered around the source of interest. The normalization of both the target and the sources included in the model are left as free parameters, while the spectral parameters of the sources in the ROI are fixed to the values from the 4FGL-DR2 catalog. Moreover, the Galactic and isotropic backgrounds are also considered in the model, with the \texttt{gll\_iem\_v07.fits} and \texttt{iso\_P8R3\_SOURCE\_V3\_v1} versions, respectively. The normalization of both diffuse components is also left free to vary during the analysis. A first iterative analysis is used aiming to maximise the likelihood between the data and the model, initially fixing the spectral parameters of the source of interest. Then a second fitting process is performed leaving the spectral parameters of the target free to iterate. A limit of Test Statistics (TS) TS=4 (approximately 2$\sigma$) is used to compute the fluxes of each bin of the LC. The integral flux is computed for each bin with TS$\geqslant$4. On the other hand, for TS<4, an upper limit of the flux is derived. The LCs available at the LCR are extracted with 3-day, 7-day and 30-day bins. Here, we use the 30-day binned LCs to introduce the minimum number of upper limits, in order to have evenly-sampled data series. In case there are still upper limits in the LCs, we treat them as non-detections and we do not include them in the periodicity analysis.

\subsection{Radio: OVRO data}
To complement the optical and $\gamma$-ray periodicity analysis we collected radio data from the 40m telescope from Owens Valley Radio Observatory (OVRO). This telescope observed the blazar sample included in this work as part of a much larger blazar monitoring program \citep{richards2011}\footnote{\url{https://www.astro.caltech.edu/ovroblazars/}} in the 15 GHz band. Here we make use of the OVRO data from January 2008 until June 2020. The data analysis procedure was performed according to the prescriptions detailed in \cite{richards2011}. We include the radio LCs of 10 out of the 15 blazars presented here, with a time span of roughly 12.5 years. To avoid introducing measurements with very large uncertainties that could lead to a poor estimation of the period, we consider those data points with a signal to noise <3 as non-detections, excluding them from the analysis. 

\section{Periodicity analysis}\label{sec3}
\subsection{Methodology and analysis tools}\label{sec3.1}
The methodology used in this work is based on two analysis steps. With the first step, due to the large amount of LCs analysed here, we make a preliminary analysis to perform a filtering of those candidates that may present a {quasi-periodic} emission, and those that do not show any hint of displaying a {quasi-periodic} modulation. This is made due to the high computational time needed to simulate and analyse some of the data sets, spanning up to more than 40 years. This first filtering is made with the first stage of the pipeline developed by \cite{penil2020}. This step uses two different methods to search for periodicities: Phase Dispersion Minimization (PDM) and REDFIT. With these methods, we select those sources showing consistent periods with significances $\gtrsim$2$\sigma$. Those sources with no signatures of periodicities and significances $<$2$\sigma$ are excluded from the analysis. For this first screening, we have not taken into account the correction of the significance due to the trial factors \citep[also known as ``look-elsewhere'' effect, see][]{gross2010}.

Then, following the methodology already used in \cite{otero-santos2020}, the final results are obtained with three different methods for the periodicity analysis: the Z-Discrete Correlation Function (ZDCF), the Lomb-Scargle periodogram (LS) and the Weighted Wavelet Z-transform (WWZ). Astronomical data series are typically irregularly sampled due to observational gaps. Thus, these methods have been optimised to have a better performance under this condition, and they have been applied in the past in several temporal studies in blazars \citep[see e.g.][]{sandrinelli2016,zhang2017a,hong2018,covino2019,penil2020,li2021}. 

\subsubsection{Phase Dispersion Minimization}
The PDM is a method based on the periodicity technique tool developed by \cite{lafler1965} to estimate periods in data from RR Lyrae stars. It evaluates the scatter of the data and its variance around the mean LC \citep{stellingwerf1978}. The data are grouped into phase bins to evaluate different periods, and the scatter w.r.t. the mean LC is estimated through a parameter $\theta$. If the evaluated period is not a true period, then $\theta\approx1$. For this method, periods appear as a local minima in the value of $\theta$. This method is based on a least-square fit relative to the mean curve, which is defined as the means of each bin. The PDM is well suited for data series with a reduced number of data points and a limited time coverage. Additionally, it performs well when analysing data series with non-sinusoidal periodicities (e.g. recurrent flares).

A proper phase binning is needed to perform a reliable analysis while avoiding unbearable computational times. Some caveats have also been evaluated for this method in the past. Due to its performance, it finds all periodic components. Thus, harmonics of the true periodic signal will appear in the PDM representation. Moreover, if different periods are present in the data set, one should remove the different oscillations in order to search for the weaker modes. A detailed discussion on the method and its caveats or restrictions are presented by \cite{stellingwerf1978}.

\subsubsection{REDFIT}
REDFIT is a software programmed in \textit{Fortran 90} and developed by \cite{schulz2002} to analyse climate time series, characterised by its sparse sampling, finite length, and dominated by red noise. Thus, it represents an adequate tool for the analysis of time series of blazars, characterised by the same limitations. It is based on a first-order autorregresive (AR1) model \citep{hasselmann1976} to test whether the variability observed in data series corresponds to true periodic oscillations or to stochastic fluctuations. The AR1 process models the variations as Gaussian (white) noise plus a term whose variability has a dependence with previous data points of the series with $\tau$, the characteristic time scale (a measurement of the ``memory'' of the model). The AR1 model that describes the data and the periodogram of the time series are calculated. Then, the periodogram peaks are evaluated by simulating time series using this AR1 model.

This method works under the assumption that the background noise is well modeled with an AR1 process. Moreover, an accurate estimation of $\tau$ is important to compute the AR1 that represents the data series. Also, the code is programmed to report significances up to a level of 2.5$\sigma$. The method and its caveats are evaluated in detail in \cite{schulz2002}.

\subsubsection{Z-Discrete Correlation Function}
The Discrete Correlation Function (DCF) is a robust method for temporal analysis. It has the advantage of not interpolating the data when dealing with unevenly sampled data series, contrary to Fourier transformed based methods \citep{edelson1989}. In particular, the ZDCF was developed to have an improved performance under these conditions \citep{alexander2013}. The different approach in the binning of the correlation function and the use of the z-transform are the main advantages w.r.t. the classic implementation of the DCF. A detailed discussion on the technical details of the ZDCF can be found in \cite{alexander2013}. Moreover, this tool has the advantage of being able to detect repeating patterns despite not being purely sinusoidal (e.g. recurrent spikes in the LC due to bright flares). When applying the ZDCF on a sinusoidal or recurrent data series, we expect to see an autocorrelation curve with a sine-like behaviour, showing repeating maxima/minima with a separation equal to the value of the period.

Among the caveats of this method, we highlight those related to phase and amplitude variations in the LC, noise and systematic effects. These effects can have a strong impact on the ZDCF, leading to features in the autocorrelation curve. Changes on the amplitude of the LC or in the phase of the modulation can provoke partial autocorrelations, attenuation of the maxima/minima, alternating high and low peaks or long term trends in the correlation curve, that may make difficult or impede the detection of the underlying periodic modulation of the data \citep[see e.g.][]{mcquillan2013}. Moreover, red noise can mimic QPOs in non-periodic data series, leading to false detections. Thus, this type of noise is the most relevant when dealing with finite and unevenly sampled data series, and must be treated carefully when analysing blazar LCs. 

\subsubsection{Lomb-Scargle periodogram}
The LS periodogram is one of the most extended tools for temporal and periodicity analysis of unevenly-sampled data series. It is based on the discrete Fourier transform, but it was modified w.r.t. the classical periodogram so that the least-square fit of the sine functions
is minimised, as well as to work without performing interpolations in these kinds of time series \citep[][]{lomb1976,scargle1982}. These modifications can help to eliminate or reduce the impact of spurious features or peaks in the periodogram. A detailed analysis on the performance of the LS compared to the classic implementation of the periodogram can be seen in \cite{vanderplas2018}. The LS periodogram is thus a powerful tool for detecting sine-like signals in irregularly sampled series. However, it may lose sensitivity when dealing with recurrent patterns that are not purely sinusoidal.

The LS periodogram is also affected by several caveats or effects that may have an impact when analysing time series. It is important to choose a proper frequency grid to sample the frequency space in order to not lose relevant spectral features while avoiding very large computational times. Moreover, spurious peaks may appear at a period $n \cdot P_{\text{true}}$, associated with harmonics of the true period. Also, false detections can be related to regular gaps in the LCs (e.g. observational gaps of astrophysical sources during periods of no visibility can introduce a fake peak at a period of 1 year). Additionally, as for the ZDCF, amplitude variations in the LC can lead to a splitting, displacement or vanishing of the true peak. A detailed discussion of some of these caveats can be found in \cite{vanderplas2018}. Moreover, red noise also has a strong impact on the LS periodogram, causing the appearance of big bumps at high period values (low frequencies) that can be wrongly associated with true periodic oscillations. Finally, trends or jumps in the LC can also lead to the wrong identification of the true period of the signal \citep[][]{mcquillan2013}.

\subsubsection{Weighted Wavelet Z-transform}
Another method derived from the Fourier transform is the wavelet transform. This tool decomposes a time series in the frequency and time domains, studying a time series in the time-frequency space \citep{torrence1998}. This allows not only to detect periodic signals but also to localise them in time. As for the ZDCF and the LS periodogram, a modification of the classic wavelet is used in this work to deal with irregularly sampled data series, the WWZ \citep{foster1996}. In particular, the WWZ uses the z-transform to improve its performance w.r.t. the classic wavelet when analysing unevenly-sampled time series \citep{witt2005}.

Similarly to the LS, for this method it is also important the choice of the frequency and time grids to have an optimal sampling of the frequency and time spaces in order to detect the true periodic signal. The choice of the wavelet function \citep[based on the Morlet wavelet, see e.g.][]{grossmann1989} can also lead to the appearance of spurious peaks in the results \citep{torrence1998}. The finite nature of the data is taken into account by the computation of the cone of influence (COI). This region is defined as the region of the wavelet outside of which the edge effects become important to the detection of a periodic signal as a result of the finite length of the time series and the red noise. Additionally, as it happens for the LS periodogram, the effect of red noise strongly affects the WWZ at low frequencies, introducing big bumps that can mimic a true periodicity.

\subsection{Estimation of the statistical significance}
In order to quantify the reliability of a detection, we need to estimate the statistical significance of a result. However, the impact of red noise in the periodicity analysis can lead to false detections that mimic the results of a true periodic signal. Thus, to compute the significance of the results taking into account the impact of the red noise accordingly, we follow the procedure described in \cite{max-moerbeck}. We simulate $10^{4}$ non-periodic time series with the same statistical properties as each of our LCs, i.e. the exact same sampling and temporal coverage, the same power spectral density (PSD) and the same probability density function (PDF). For this, we use the \textit{python} package \texttt{DELCgen} developed by \cite{connolly2015}, implementing the procedure described in \cite{emmanoulopoulos2013}. This method is based on the LC simulation procedure proposed by \cite{timmer1995} (hereafter TK) and used in previous works, improving the statistical resemblance of the simulated LCs w.r.t. the real data series.

For the ZDCF, we now apply the methodology described by \cite{mcquillan2013}. After simulating $10^{4}$ artificial LCs, and assuming a normal distribution of the autocorrelation value of each bin of the DCF curve, we compute the autocorrelation of each non-periodic LC and the statistical significance is derived using this distribution. We estimate the significance intervals from 1$\sigma$ to 5$\sigma$. Finally, in the case of a positive detection of a periodic signal with this method, the uncertainty of the derived period is estimated using Equation (3) from \cite{mcquillan2013}.

Alternatively, for the LS periodogram we follow the procedure described by \cite{vaughan2005}. According to this method, the PSD of a pure red noise signal can be modeled as a power law of the form $P(f)=Nf^{\alpha}$, with a spectral index $\alpha=-2$. Thus, we can estimate the red noise continuum of the PSD by fitting the resulting spectral density to a power law (or equivalently, the logarithm of the PSD to a linear function w.r.t. the logarithm of the frequency). With this estimation of the continuum, we compute the different confidence levels, which will have a power-law shape to account for the increasing impact of the red noise at higher periods (lower frequencies). In order to calculate a rough estimation of the uncertainty of the derived period value, we use the half width at half maximum (HWHM) of the peak, following the procedure of previous studies \citep[see e.g.][]{bhatta2016,zhang2017a,zhang2017b}.

Concerning the WWZ, the significance is computed similarly to the LS. For each artificial LC, the WWZ and the corresponding PSD are calculated. Then we compute the mean PSD of the simulated LCs. Finally, the different confidence intervals are estimated from the mean PSD of the non-periodic time series, affected by approximately the same red noise as our data. For this calculation we follow the procedure from \cite{vaughan2005}. The uncertainty in the determination of the period value is obtained from the HWHM of the corresponding peak as for the LS periodogram. 

The final results presented in this work take into account the number of independent frequencies scanned in the analysis, as well as the number of sources analysed, in order to correct the local significance values for the trial factors \citep{gross2010}. Thus, the reported significance for the results found here corresponds to global significance, i.e. after the trial correction. This global significance is commonly used in periodicity analyses to account for the ``look-elsewhere'' effect \citep[see e.g.][]{vaughan2005,vaughan2010,sandrinelli2018}. The correction is applied as follows,
\begin{equation}
p_{global}=1-(1-p_{local})^{N},
\label{global_significance}
\end{equation}
where $N$ is the total number of trials considered. $N$ is estimated through Monte Carlo simulations \citep[see for instance][]{cumming1999} taking into account the total number of sampled frequencies (typically 100-150 depending on the time span of the data), following the approach from \cite{penil2022}. For the present case, a total of 30 independent frequencies shows a good agreement with our data.

\subsubsection{PSD fit}
\begin{figure}
	\includegraphics[width=\columnwidth]{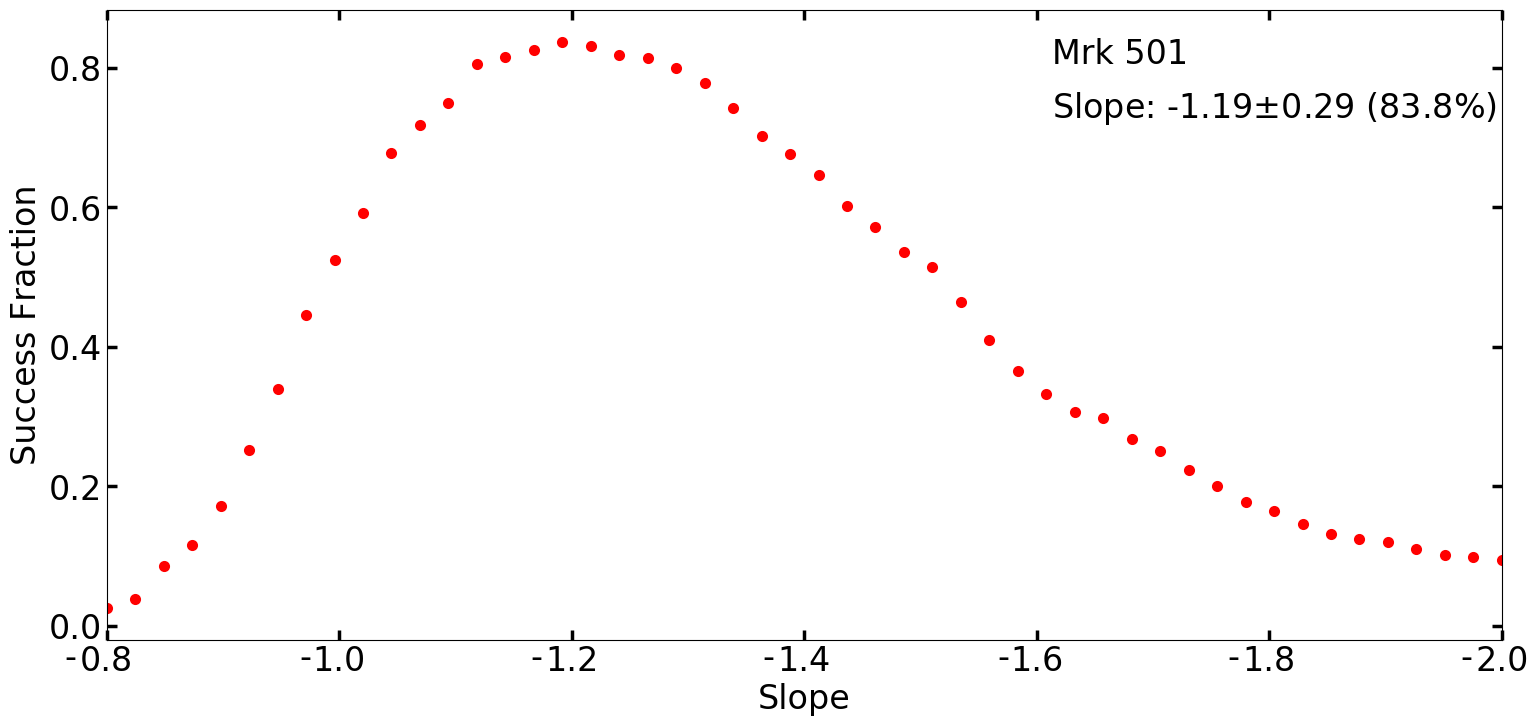}
    \caption{Example of the derived PSD slope using the PSRESP method.}
    \label{psresp_example}
\end{figure}

The significance estimation involves the simulation of artificial LCs with the same PSD as the real data. To estimate the PSD of our data and extract accurate parameters that fit this PSD, we make use of the power spectral response method implemented in the \textit{python} package \texttt{PSRESP}\footnote{\url{https://github.com/wegenmat-privat/psresp}}. This algorithm uses a Monte Carlo approach, taking into account gaps and the binning of the data, fitting the underlying PSD with a simple power law function to estimate its spectral index. It uses the LC simulation method to obtain the derived slope of the PSD. A grid of trial slopes is evaluated and the goodness of the fit is calculated with a chi-squared statistic. The error on the derived slope value is determined as the HWHM of the slope distribution. An example of the method is shown in Figure \ref{psresp_example}.

\subsubsection{Number of false detections}
\begin{figure*}
\centering
\subfigure{\includegraphics[width=0.49\textwidth]{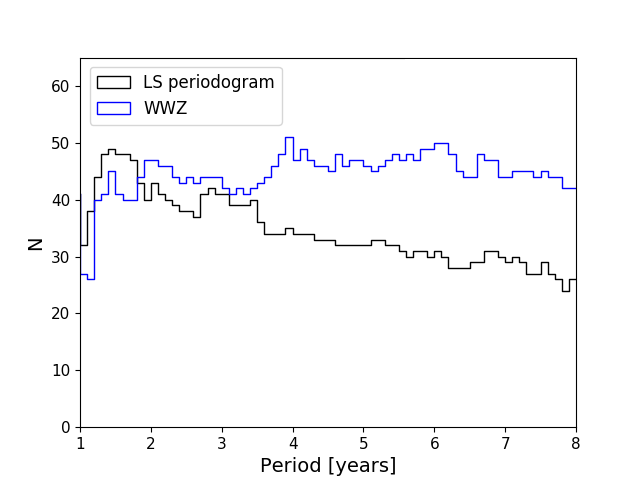}}
\subfigure{\includegraphics[width=0.49\textwidth]{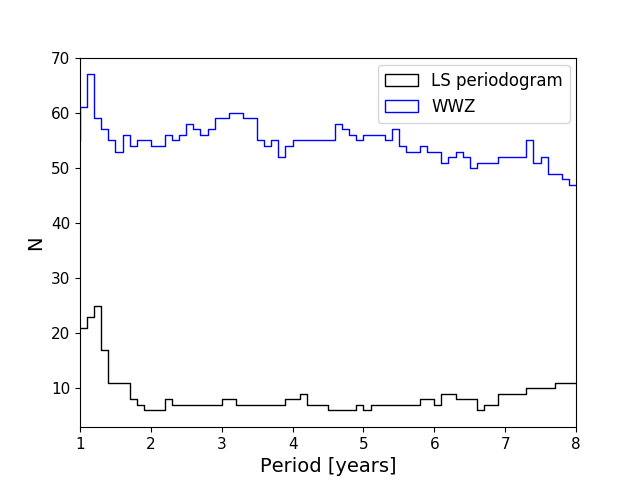}}
\caption{False detections of non-periodic signals in 10000 artificial TK time series. \textit{Left:} LS periodogram (black) and WWZ (blue) at a 5$\sigma$ confidence level for Gaussian time series. \textit{Right:} LS periodogram (black) and WWZ (blue) at a 5$\sigma$ confidence level for log-normal time series.}
\label{false_detection_rate}
\end{figure*}

The simulation of statistically identical artificial LCs to our data is a powerful tool for estimating the significance of the results, since the simulated time series would be affected by red noise in the same way as the real data. However, this red noise can still lead to false periodicity detections in non-periodic data series. To study the probability of false detections, we estimate the number of artificial and non-periodic LCs that are reporting by chance, due to stochastic variations caused by the red noise, a positive peak in the LS periodogram used in this work. For this purpose, we generate 10000 non-periodic LCs with the TK algorithm. For this simulation, we use the same sampling as a real LC (for our case, we have used the sampling and time coverage of the R-band LC of Mrk~501, spanning $\sim$19 years), and a PSD index $\alpha=-2$. This higher spectral index w.r.t. the LCs analysed in this work ensures that the estimated false positive rate derived from this test is an upper limit of the false positive rate of the real data sets. This leads to false detection rates ranging between 0.3-0.5\% above a significance of 5$\sigma$ for periods between 1.0~years and 8.0~years (i.e. 1 false detection per 200 analyses). The results are shown in Figure \ref{false_detection_rate}. We conclude that despite the low false detection rate, the TK algorithm leads to a higher rate than expected for each significance level. The same experiment was repeated for the WWZ, finding a similar false detection rate, as shown in the left panel of Figure \ref{false_detection_rate}.

It is known that real data series from blazars are better described by a log-normal distribution \citep[see e.g.][]{giebels2009,sinha2018,shah2018}. However, the TK algorithm is based on a gaussian-like LC generation, most likely leading to an overestimation of the statistical significance. \cite{shah2020} propose a conversion of a normal into a log-normal data series through the exponential of the former. Therefore, we also test this hypothesis with the exponential TK LCs. The false detection rate estimated for the LS periodogram of the log-normal artificial data series was found to be relatively similar to that from gaussian TK time series at significances of 1$\sigma$-2$\sigma$. For higher confidence levels, the false detection rate falls to 0.1\% at 5$\sigma$. On the other hand, the rate derived with the WWZ does not show any significant improvement, even for the 5$\sigma$ confidence level, with a false detection rate of $\sim$0.5-0.6\%, as observed in Figure \ref{false_detection_rate}. These results are still higher than the expected rate for the corresponding confidence levels. This test proves that the TK LCs can provide a fast and fairly good approach, but they tend to overestimate the resulting significances.

\begin{figure}
\centering
\includegraphics[width=\columnwidth]{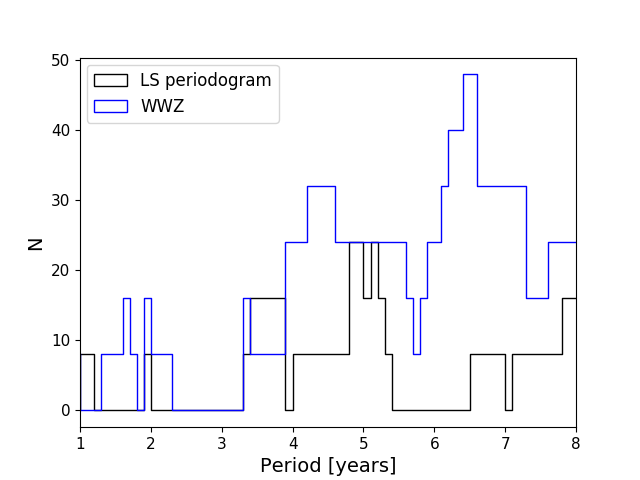}
\caption{False detections of non-periodic signals in 10000 artificial time series generated with the approach from \protect\cite{emmanoulopoulos2013} at a confidence level of 3$\sigma$. The black and blue distributions correspond to the false detections measured with the LS and the WWZ, respectively.}
\label{false_detection_rate_emm}
\end{figure}

Finally, the performance test of the significance estimation was extended to the artificial LCs generated with the method from \cite{emmanoulopoulos2013}, that gives a more rigorous treatment to the red noise. The results for this case are presented in Figure \ref{false_detection_rate_emm}. The false detection rates for the LS and WWZ range between $\sim$20-30\% for a 1$\sigma$ confidence level, and $\sim$0.24-0.32\% for a 3$\sigma$ significance (1 false detection per $\sim$300-400 data series at a 3$\sigma$ level). This is in line with the expected probabilities for each confidence level (68.27\% and 99.73\%, respectively). For a higher significance, no false detections are observed. This is expected given the ``reduced'' number of simulations performed for this test, and the corresponding probabilities in the case of 4$\sigma$ and 5$\sigma$ confidence levels. However, this is a good indication that the significance estimation is reliable when using the simulation approach from \cite{emmanoulopoulos2013}.

\subsection{Performance tests}

\begin{figure*}
\centering
\subfigure{\includegraphics[width=0.33\textwidth]{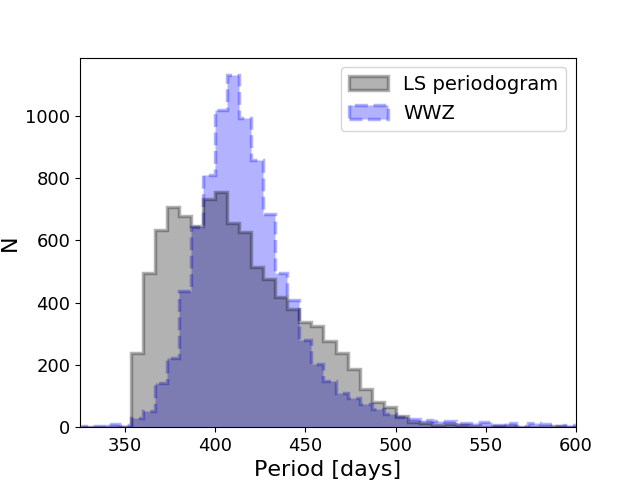}}
\subfigure{\includegraphics[width=0.33\textwidth]{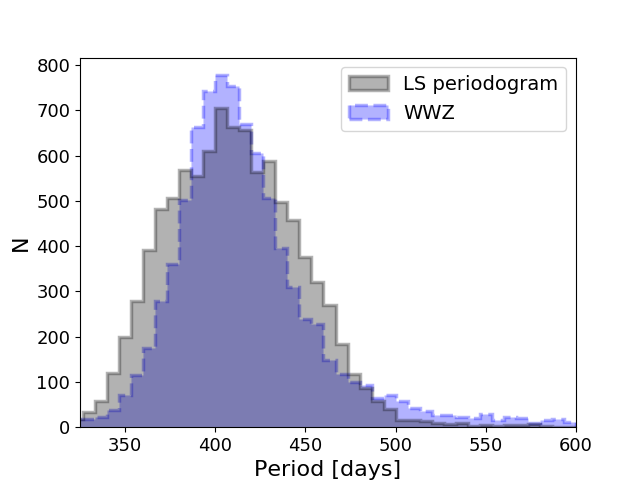}}
\subfigure{\includegraphics[width=0.33\textwidth]{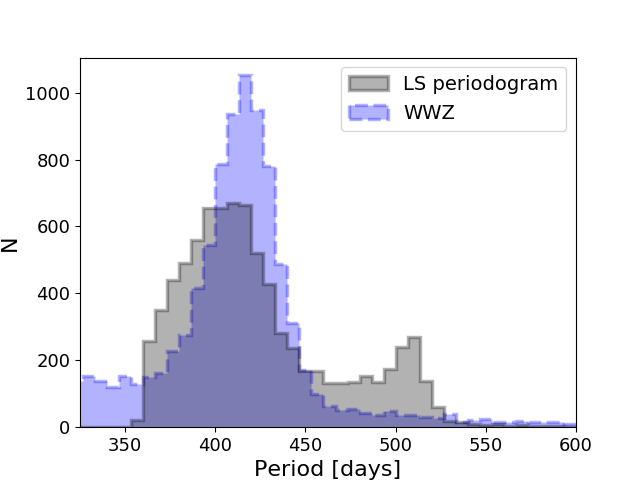}}
\caption{Distribution of the measured period for the different performance tests. \textit{Left:} 400-day periodic LC with a red noise signal. \textit{Middle:} 400-day periodic LC with a red noise signal and 20-day binning. \textit{Right:} 400-day periodic LC with a red noise signal after removing 10\% of the data set. The black and blue distributions correspond to the LS and WWZ, respectively.}
\label{performance_tests}
\end{figure*}

The analysis tools explained before were selected due to their increased performance under realistic conditions when analysing unevenly-sampled data. In order to understand the impact of these conditions, we performed different tests to assess the impact of the red noise, the binning of the data, and the uneven sampling, in the results of the different methods. We perform these tests for the LS periodogram and WWZ methods used here, evaluating their performance analysing a known periodic signal under realistic conditions. We note that this test has not taken into account the correction on the significance by the trial factors. Therefore, all the results here refer to the local significance. Also, due to the computational burden of the tests, the simulations have been performed using the TK approach, that tends to slightly overestimate the significance of the results. However, it can give a good estimation of the effects of realistic conditions affecting real astronomical data series. These effects are already being accounted for in the simulations performed in our analysis, as the simulation method employed here uses the same binning, sampling and statistical properties of the original LCs. Additionally, these tests can provide a rough estimation of the errors expected in the period determination.

\subsubsection{Effect of the red noise}\label{sec3.3.1}
First, we evaluated the impact of pure red noise in the detection of the true periodicity. For this, we simulated a purely sinusoidal signal, with a period of 400 days and a time coverage of 1000 days. These numbers were chosen to produce LCs with a period long enough to be located in a position of the frequency space already affected by the effect of red noise, more important at lower frequencies (longer periods). We then generated 10000 artificial LCs of pure red noise, with a spectral index between -1.8 and -2.2, to contaminate the periodic signal. 

We find that after adding a red-noise signal to a true periodic data series, the LS periodogram is still able to detect $\sim$95\% of the periodicities above a significance of 5$\sigma$. However, a median shift in the measured period of $\sim$6\% of the true value is observed, with a median deviation from the nominal period of 24~days. Finally, the power ratio between the peak of the pure sinusoidal and the contaminated signals decreases by a factor 3, approximately. The left panel of Figure \ref{performance_tests} shows the distribution of measured periods after contaminating the signal with the red noise data series.

Concerning the WWZ, this method allows us to detect $\sim$91\% of the simulated periodic signals at a 5$\sigma$ confidence level after adding a red noise data series. As observed from the left panel of Figure \ref{performance_tests}, it shows a narrower distribution of the measured period than the LS periodogram. The median value of this distribution shows a smaller shift of the measured period, with a mean displacement of $\sim$18~days, approximately 4.5\% of the real value. 

\subsubsection{Effect of the binning}
The effect of binning the data has also been tested. For this, we applied a binning of 10~days, 20~days and 30~days to the periodic signal after being contaminated by the artificially generated red noise. The effect of the binning in the LS periodogram causes the loss of $\sim$8\% of the detections at a 5$\sigma$ confidence level for the 10-day binning. For larger bins sizes, this value falls to $\sim$80\% and $\sim$67\% of the detections at 5$\sigma$ for 20-day and 30-day bins, respectively. The shift on the period value is roughly the same as that for the previous test, with a median value of $\sim$26.5~days. However, the ratio between the power of the pure periodic signal and the measured contaminated peak slightly improves to $\sim$0.38 for the larger binnings due to the decrease of fast fluctuations after binning the data. The distribution of periods for the 20-day bin width is shown as an example in the middle panel of Figure \ref{performance_tests}.

On the other hand, the number of detections with the WWZ was found to be much more affected by a binning of the data than the LS periodogram. For the smaller bin tested here (10 days), the combined effect of the red noise and the binning prevents us from detecting at a level of 5$\sigma$ a total of $\sim$38\% of the simulated periodic signals. The detection rate falls to a $\sim$15\% for a 30-day binning. Hence, the WWZ has a much worse performance under red-noise affected and binned data than the LS periodogram. Moreover, the detected periods suffer a mean shift of the measured value of $\sim$21~days ($\sim$5\%). Again, this is consistent with the narrower distribution of periods observed for the WWZ. In the middle panel of Figure \ref{performance_tests} we show as an example the distribution of measured periods to show how this value is affected by the red noise and the binning, for a bin width of 20~days.

\subsubsection{Effect of the uneven sampling}

\begin{figure*}
	\includegraphics[width=0.835\textwidth]{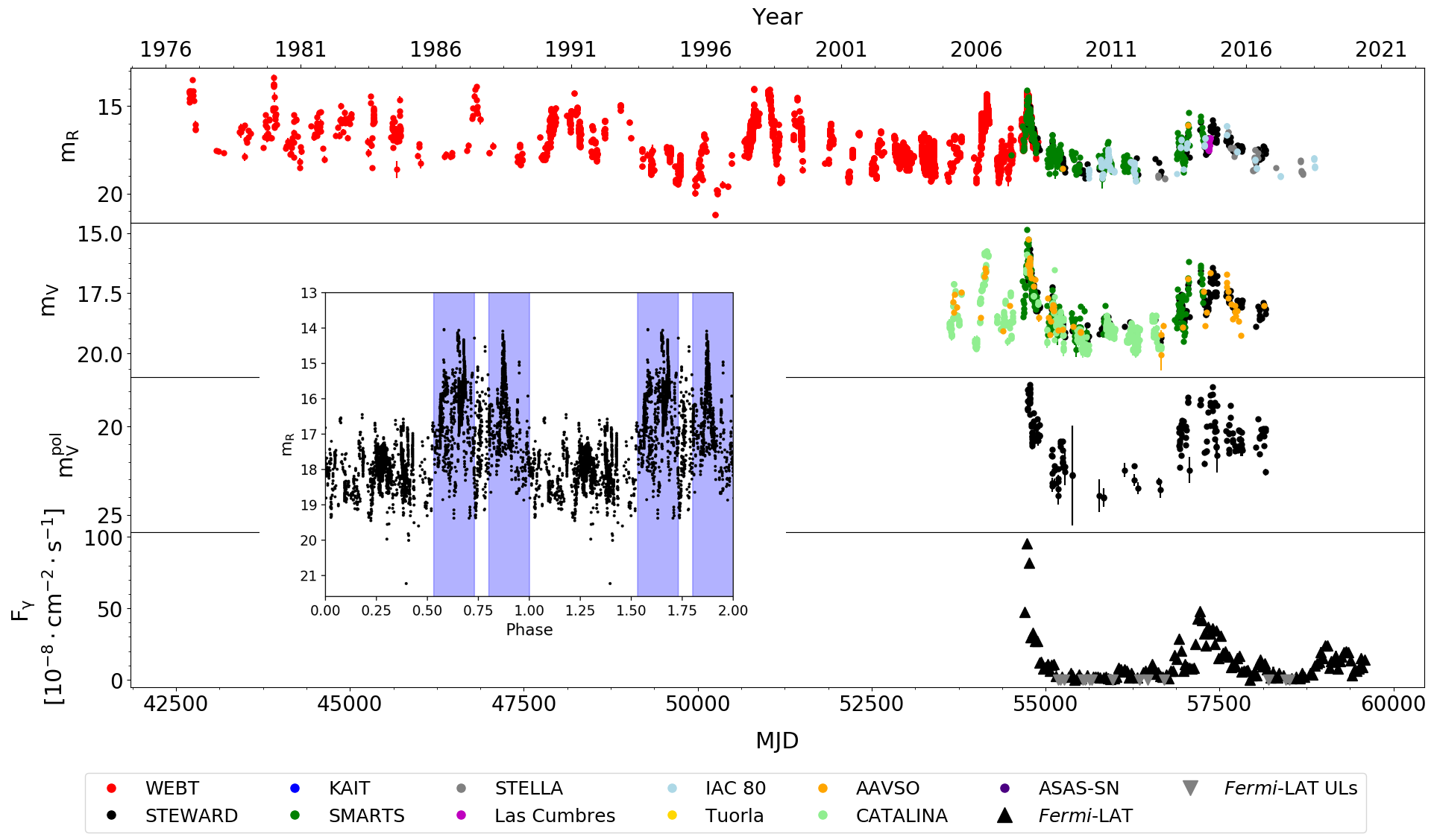}
    \caption{MWL LCs of AO~0235+164. From top to bottom: Optical R band, optical V band, optical polarization and HE $\gamma$-ray band. The inner panel shows the folded R-band LC with the 8.2-year period measured here. The blue contours show the double-peak structure of the flares.}
    \label{ao0235_lcs}
\end{figure*}

The last test performed consists of randomly removing different percentages of points from the simulated LCs to mimic the uneven sampling present in ground-based astronomical data series. We evaluate the performance of these two methods after removing between 10\% and 50\% of the data points, and adding a pure red noise signal. The LS periodogram is clearly affected by the uneven sampling even after removing only 10\% of the data, with only the 78\% of the simulations detected at a 5$\sigma$ confidence level. This value drastically affects the performance of the periodogram, which is only able to detect $\sim$42\% of the simulations after eliminating 50\% of the data points. Similar values were also reported by \cite{penil2020} for the LS periodogram of unevenly-sampled data. The shift in the period, as observed in the right panel of Figure \ref{performance_tests} where the period distribution is presented for a LC with 90\% of the original data points, is only slightly affected, showing a median change of $\sim$23~days. However, the distribution shows a broad peak, as well as a secondary maximum at $\sim$500~days, leading to a median derived period of 424~days. This indicates that the presence of gaps in the time series tends to induce the appearance of spurious peaks in the LS periodogram that can impede the detection of the true period. Finally, the ratio of the sinusoidal and simulated LC power peaks was estimated to be $\sim$0.2-0.3.

The results obtained by the WWZ report a loss of 25\% of the detections at a level of 5$\sigma$ after eliminating 10\% of the LC points, similarly to the LS periodogram. When increasing the percentage of gaps up to 50\%, the detection rate drops down to 70\%. Therefore, even if the WWZ is highly affected by red noise and gaps, it shows an overall better performance than the LS when dealing with unevenly sampled data series. The measured period reveals a mean deviation of $\sim$24~days (6\%) for a loss of 10\% of the data. This can be seen in the right panel of Figure \ref{performance_tests}. This value becomes even higher for larger gaps in the data, reaching a shift of 13\%, i.e. $\sim$53~days, after removing 50\% of the data points. As for the two previous tests, the distribution of measured periods with the WWZ presents a much narrower shape. In addition, we do not see a secondary peak at 500~days. 

\section{Results}\label{sec4}
We have analysed 15 blazars searching for signatures of {quasi-periodic} emission in their radio, optical, polarized and $\gamma$-ray emission. After the first step of the analysis (as detailed in Section \ref{sec3.1}), the following candidates did not show any hint of {quasi-periodic} modulation on their emission above 2$\sigma$ without considering the trial factors: 1ES~1959+650, CTA~102, Mrk~421, OJ~248, OJ~287 and W~Comae. The other 9 blazars showed some hint of {quasi-periodic} emission in at least one of the MWL LCs presented in this work with a local significance $\geqslant$2$\sigma$. Thus, the detailed analysis was used to evaluate the periodicity hypothesis in their emission. This threshold is chosen as a typical significance value of reported periodicity hints in the literature \citep[see e.g.][]{sandrinelli2018,penil2020}. The results of each source are presented in the following subsections. Moreover, the inferred periods and their significance are summarised in Table \ref{periodicity_results}. The results of the band with the highest global significance are presented in this section in Figures \ref{ao0235_results}, \ref{pks1222_results_pol}, \ref{mrk501_results}, \ref{bllac_results_pol} and \ref{1es2344_results} for the sources reaching a post-trial significance $\geqslant$2$\sigma$. The figures for the rest of the bands for each source, as well as an example of those sources with negative results, are presented as supplementary online material (see Figures \ref{pks1222_results} to \ref{3c454_3_results}).

\subsection{AO 0235+164}
\begin{figure*}
\centering
\subfigure{\includegraphics[width=0.85\columnwidth]{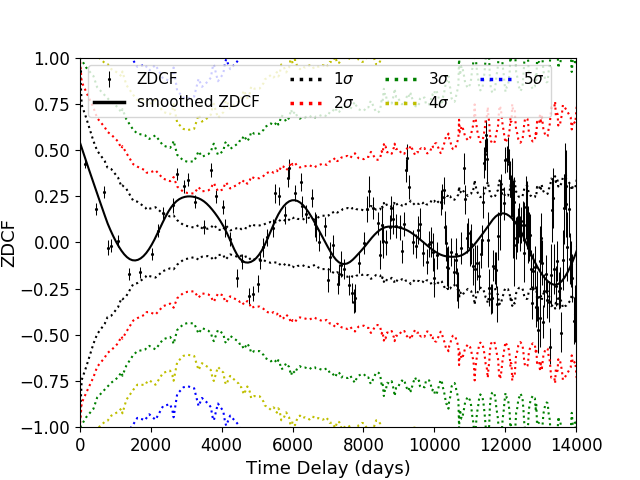}}
\subfigure{\includegraphics[width=0.85\columnwidth]{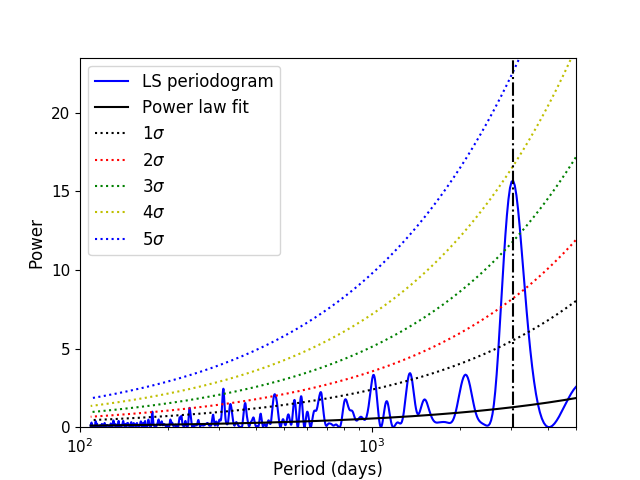}}
\subfigure{\includegraphics[width=0.85\textwidth]{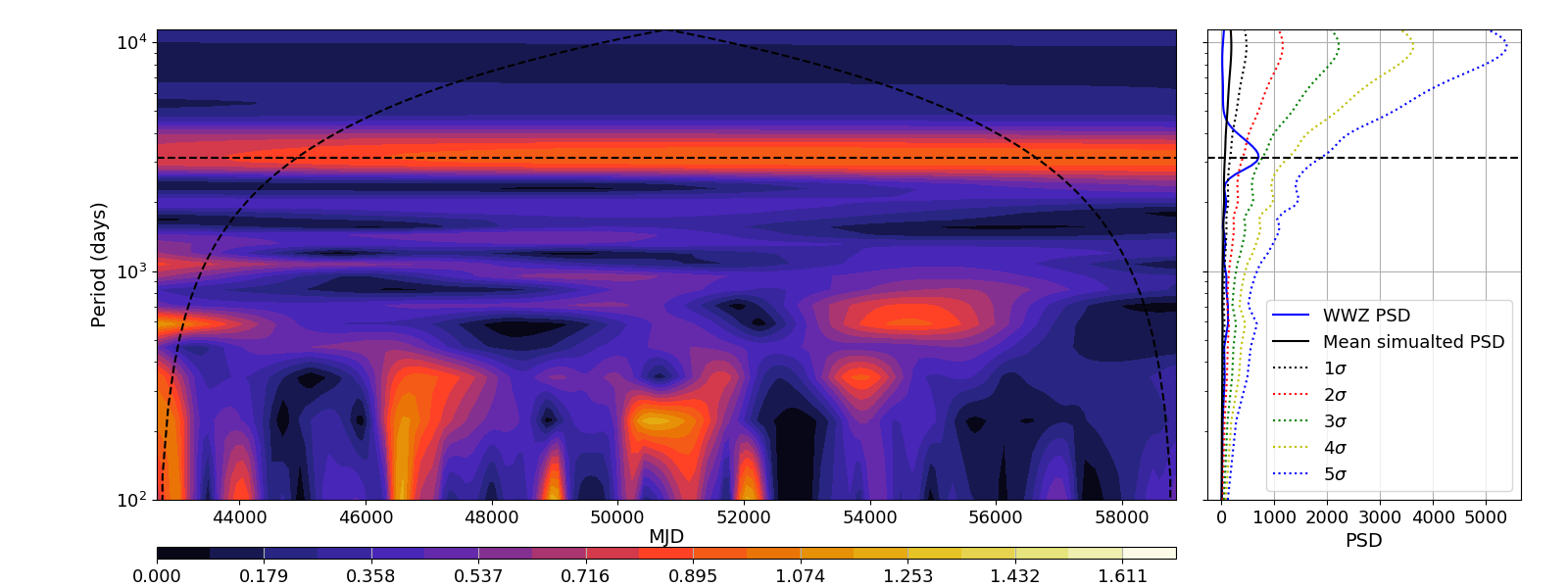}}
\caption{Periodicity analysis of the R-band data of AO~0235+164. \textit{Top left}: ZDCF method. \textit{Top right}: Lomb-Scargle periodogram. \textit{Bottom}: WWZ method. The left panel shows the power spectrum as a function of period and time. The black dashed curve represents the COI. The right panel shows the PSD. The black horizontal dashed line marks the peak of the PSD.} 
\label{ao0235_results}
\end{figure*}

The {QPO} search performed for AO~0235+164 reveals a possible {quasi-periodic} modulation in its R-band LC with a period of 8.2~years. The ZDCF has a clear sinusoidal shape with roughly equidistant maxima/minima compatible with a period of $8.2 \pm 0.3$~years, at confidence levels around 2.0$\sigma$. Moreover, the LS periodogram displays a power peak at a period of $8.3 \pm 0.6$~years and a significance of 3.8$\sigma$. Similarly, the WWZ presents a peak extending throughout all the monitored period at $8.7 \pm 1.5$~years and a statistical significance of 2.9$\sigma$. These results can be observed in Figure \ref{ao0235_results}. This period is compatible with that found by \cite{raiteri2006} and recently by \cite{roy2022}. 

Given the large value of the period found in the optical R-band LC, we cannot test this {QPO} in the other bands due to their much shorter time coverage ($\lesssim$13~years) w.r.t. the former. However, by visual inspection of Figure \ref{ao0235_lcs}, we observe that two of the flares visible in the R-band are also present in the polarization, $\gamma$-ray and radio data series occurring simultaneously, indicating that this period could be present in the MWL emission of AO~0235+164. Nevertheless, a much larger data set is needed in these bands to confirm the presence of this modulation. Assuming the aforementioned {QPO}, the next outburst would occur at the beginning of 2024. The observation of this event would help to confirm the existence of this periodicity.

\subsection{PKS 1222+216}
This FSRQ presents a possible {QPO} in the PDM and REDFIT methods used in the first analysis filter with a value of $\sim$1.2~years. This period is observable in the R- and V-band LCs, but appears especially strong in the polarized emission. The dedicated analysis based on the simulated data series agrees with the previous measurement in the polarized LC of PKS~1222+216. These results are presented in Figure \ref{pks1222_results_pol}. The LS periodogram and the WWZ present a power peak at a period of $1.2 \pm 0.1$~years and $1.3 \pm 0.1$~years, with significances of 3.3$\sigma$ and 3.0$\sigma$, respectively. On the other hand, the shape of the ZDCF is dominated by the two brightest flares (at approximately MJD~55500 and MJD~57000) with respect the rest of the maxima in the LC. However, an underlying faster oscillation of the correlation coefficient compatible with the observed period ($1.1 \pm 0.1$~years) can be seen especially in the first half of the ZDCF curve. Therefore, the ZDCF of this data series shows the effect of attenuated maxima and minima commented in Section \ref{sec3.1}.

\begin{figure*}
	\includegraphics[width=0.815\textwidth]{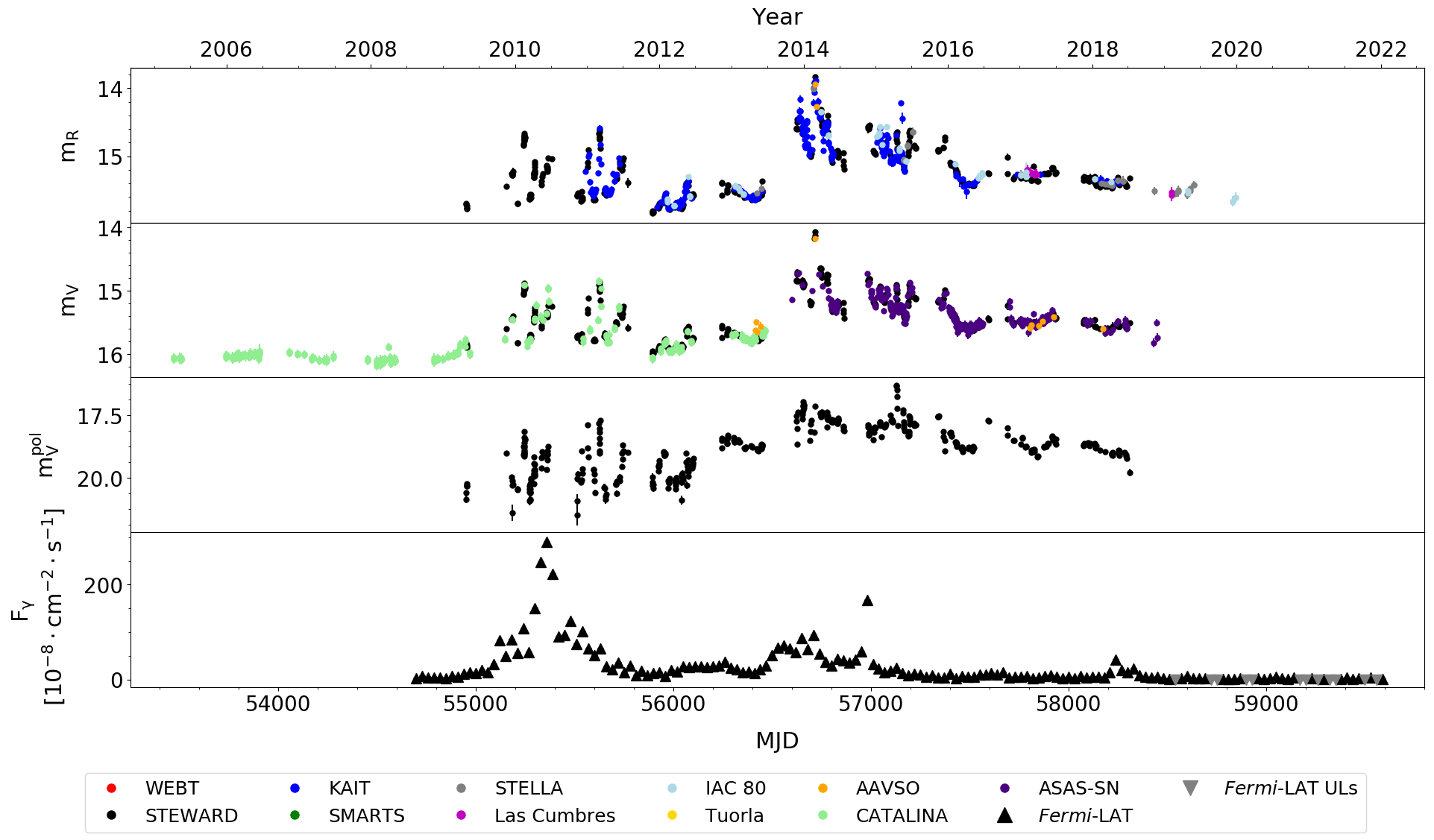}
    \caption{MWL LCs of PKS~1222+216. From top to bottom: Optical R band, optical V band, optical polarization and HE $\gamma$-ray band.}
    \label{pks1222_lcs}
\end{figure*}

\begin{figure*}
\centering
\subfigure{\includegraphics[width=0.85\columnwidth]{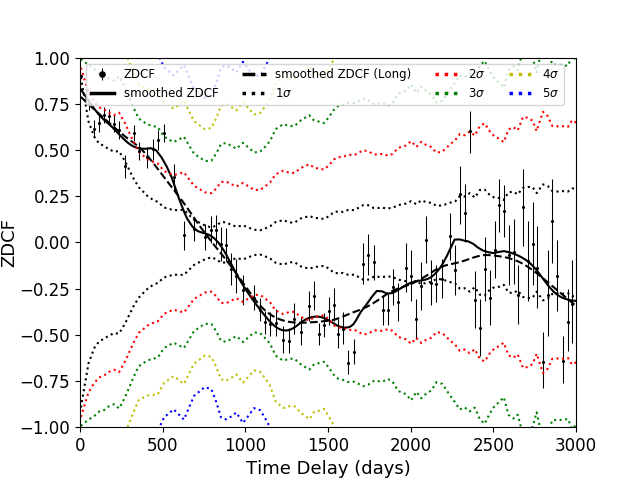}}
\subfigure{\includegraphics[width=0.85\columnwidth]{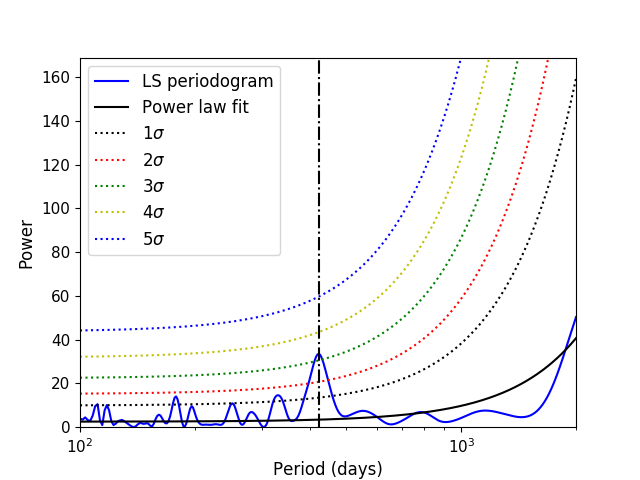}}
\subfigure{\includegraphics[width=0.85\textwidth]{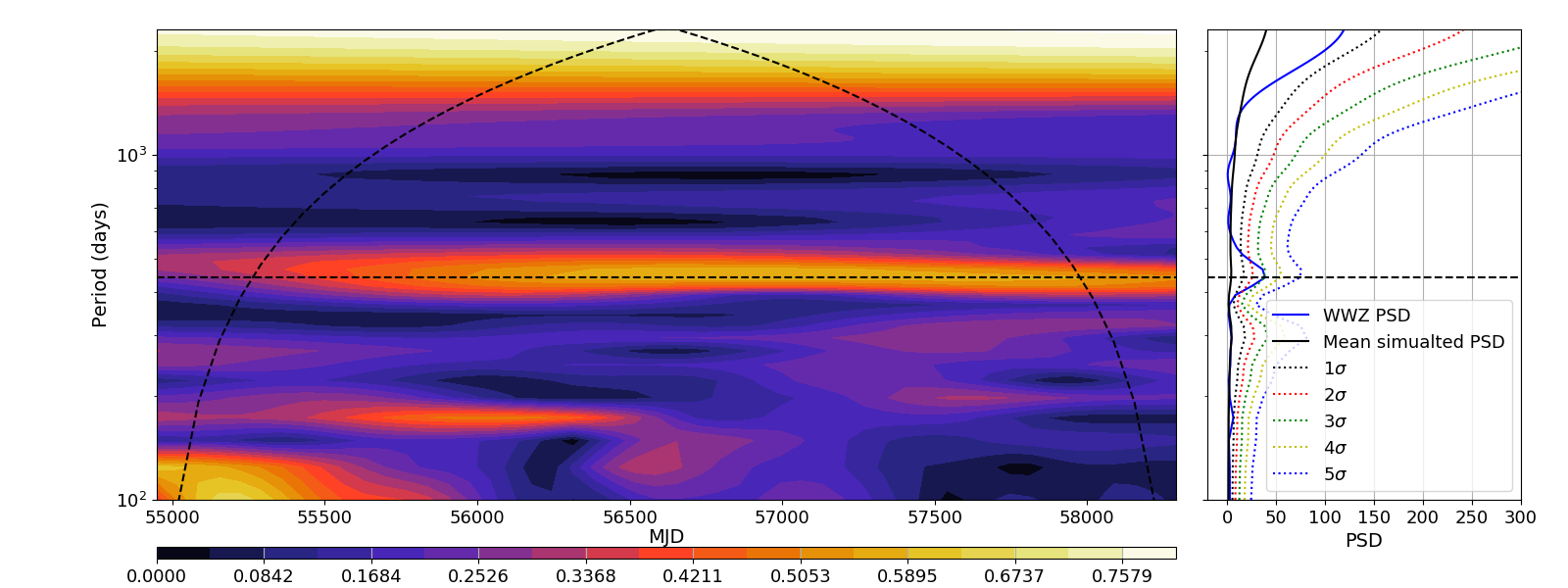}}
\caption{Periodicity analysis of the optical polarized data for PKS~1222+216. Same description as Figure \ref{ao0235_results}.} 
\label{pks1222_results_pol}
\end{figure*}

The total R- and V-band LCs show a compatible hint of {quasi-periodic} emission, with values reported by the LS periodogram and the WWZ of $1.3 \pm 0.2$~years. However, the significance of these results is lower than that from the polarized emission, with values $\lesssim$2$\sigma$, as reported in Table \ref{periodicity_results}. Moreover, the ZDCF of the total light shows the same behaviour as the one observed for the polarized flux, with a slower modulation of $\sim$1500-2000~days, and the underlying faint oscillation compatible with the potential periodicity, with a value of $1.2 \pm 0.1$~years for both bands (see Figure \ref{pks1222_results}). 

The HE $\gamma$-ray LC presents a possible preliminary {QPO} of $\sim$4~years. This is caused by the same long modulation observed in the optical emission, introduced by the bright flares at approximately MJD~55400, MJD~56700 and MJD~58200, respectively, as shown in Figure \ref{pks1222_lcs}. 
The ZDCF shows three maxima and minima with a distance of 4.0~years. Due to the different intensity of the flares, these peaks in the ZDCF show alternating maxima and minima. This is especially noticeable for the minima of the correlation curve. Moreover, the LS periodogram and the WWZ also reflect the presence of this modulation, at periods of 4.1~years (0.9$\sigma$) and 3.8~years (1.0$\sigma$), respectively. Given the low significance of the results derived for the $\gamma$-ray emission of PKS~1222+216, the red noise provides a plausible explanation to this behaviour. Finally, no radio data are available for this blazar in order to test the observed period.

\subsection{3C 273}
The analysis reports a negative result for the optical R- and V-band LCs, as well as the \textit{Fermi}-LAT $\gamma$-ray data set. A peak in both the LS periodogram and the WWZ appears in all three data sets with values between 6.0~years and 8.2~years, however not statistically significant and most likely caused by the influence of the red noise, as derived from the analysis (see Figure \ref{3c273_results}). Moreover, the almost unpolarized emission presented by this blazar also leads to no significant period detected in its polarization degree. \cite{fan2001} report a period of 13.65~years in the optical B-band emission of this blazar in a 110~year LC. However, due to the much shorter coverage of our data sample, we cannot aim to prove such a long period.

On the other hand, the radio LC presents a peak above 2$\sigma$ in both of the methods used in the first analysis step (PDF and REDFIT) at a period of $\sim$3~years. However, this period does not appear significant after a careful treatment of the red noise and the variability through the LC simulation used for the analysis. In addition, \cite{ciaramella2004} report a period in the radio emission of 3C~273 of 8.55~years. Again, as for the 13.65-year optical period, we cannot evaluate the existence of this {QPO} as our radio LC spans only $\sim$12~years, less than twice the period reported by the authors.

\subsection{3C 279}
The preliminary analysis shows a possible {QPO} of $\sim$4.2~years in the HE $\gamma$-ray emission of 3C~279 over a local significance of 2$\sigma$ with both employed methods. While this period is reflected in the LS periodogram and the WWZ as a power peak, the {quasi-periodicity} analysis is not able to significantly detect it over the modeled red noise with the artificial LCs. 
In order to check this period in the rest of the bands included in this work, we have carried out the {quasi-periodicity} search in the optical total and polarized light and the radio emission of 3C~279, with the same results as the $\gamma$-ray LC, as presented in Figure \ref{3c279_results}. This source has been claimed in the past to show a quasi-periodic behaviour in the optical band, with periods of 256~days and 931~days and significances <3$\sigma$ by \cite{sandrinelli2016}. However, the results reported by these authors do not consider the correction of the local significance introduced by the trial factors.

\subsection{PKS 1510-089}
A visual inspection of the broadband LCs of this blazar (presented in Figure \ref{pks1510lc}) shows that the variability of its emission is mainly dominated by intense and very fast flares. The presence of these features along the LC yields several potential {QPOs} in the first step of the analysis, performed with the PDM and the REDFIT implementations, the most promising located at a period of 1.2~years in the optical R and V bands and the \textit{Fermi}-LAT LC. This value is consistent with the one claimed by \cite{sandrinelli2016}.

After performing a rigorous analysis and taking into account the different trial factors, we cannot rule out the possibility of these periods being caused by stochastic flare-like variability (see e.g. the R-band results presented in Figure \ref{pks1510_results}). The WWZ time-frequency PSD shows a possible periodicity at $\sim$1.3~years, however with a gap present in the time-decomposition between MJD~55500-57000, approximately. This could indicate that the appearance of this signal could be due to the presence of the aforementioned flares. Moreover, the ZDCF shows a rather sinusoidal shape, with low significance (1$\sigma$-1.5$\sigma$ after trials) and some of the peaks being diluted w.r.t. the overall evolution of the correlation function. Therefore, we cannot extract a firm conclusion on the existence of this {quasi-periodic} variability.

\subsection{Mrk 501}
The BL Lac object Mrk~501 is one of the most promising candidates found by this work, with a {quasi-periodic} modulation of its optical emission of $\sim$5~years. All three methods report a compatible value of $\sim$5~years for the R-band emission, with significances $\sim$3$\sigma$. For the V-band, a value of 4.8~years is estimated from the ZDCF, LS periodogram and WWZ, with slightly lower significances than the R-band ZDCF and WWZ, but higher for the LS periodogram.
The results are presented in Figure \ref{mrk501_results}. Moreover, all the period and significance values are shown in Table \ref{periodicity_results}. This is the first report of such a period for the optical emission of Mrk~501 in the optical band. \cite{yang2007} reported in the past a {QPO} of 10.06~years that could be compatible with a harmonic of the true period found here. We extended the analysis to the polarized emission, with no significant period detected. Two peaks appear at periods of $2.1 \pm 0.2$~years and $4.2 \pm 0.7$~years in both the LS periodogram and WWZ. However, the analysis reports a significance <1$\sigma$. In addition, the ZDCF does not show the same periodic structure found for the total optical light.

\begin{figure*}
	\includegraphics[width=0.825\textwidth]{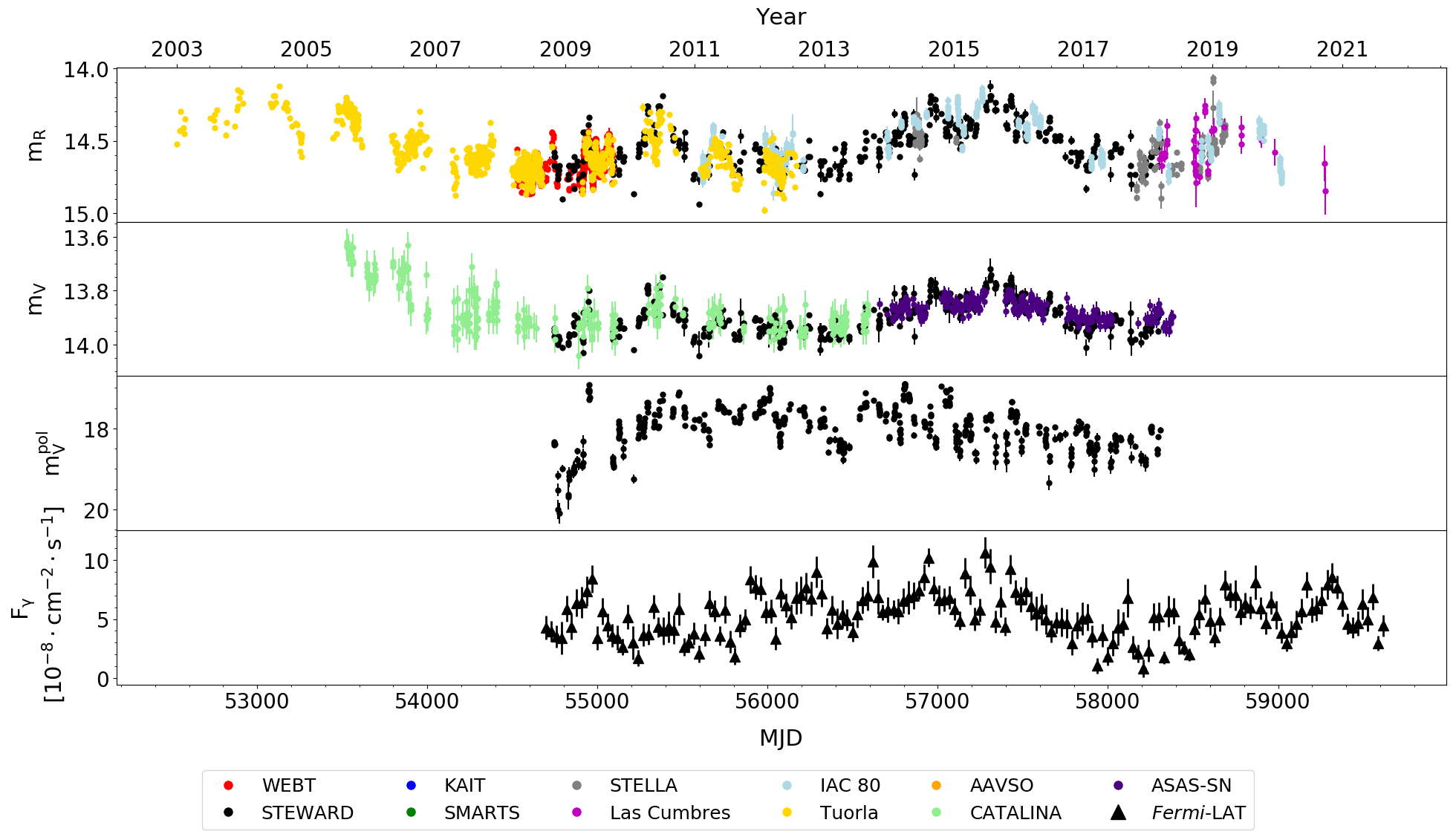}
    \caption{MWL LCs of Mrk 501. Same description as Figure \ref{pks1222_lcs}.}
    \label{mrk501_lcs}
\end{figure*}

\begin{figure*}
\centering
\vspace{-4mm}
\subfigure{\includegraphics[width=0.84\columnwidth]{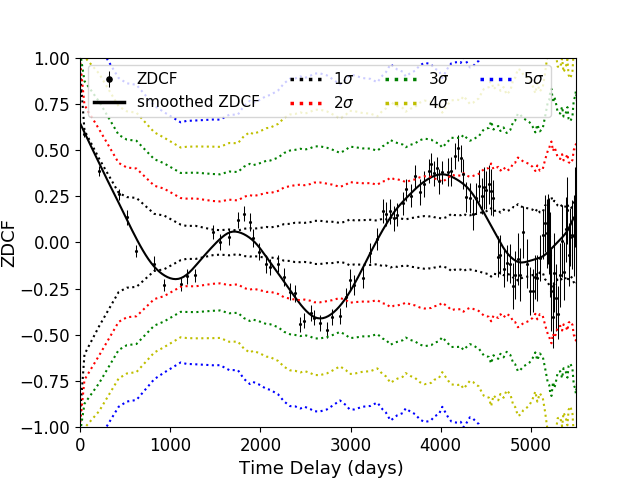}}
\vspace{-4mm}
\subfigure{\includegraphics[width=0.84\columnwidth]{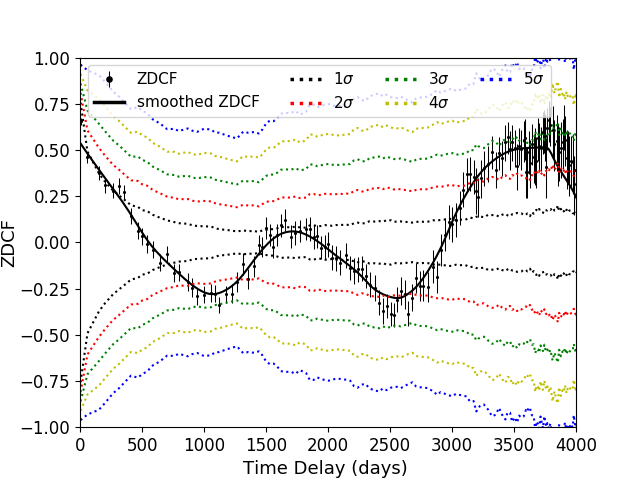}}
\subfigure{\includegraphics[width=0.84\columnwidth]{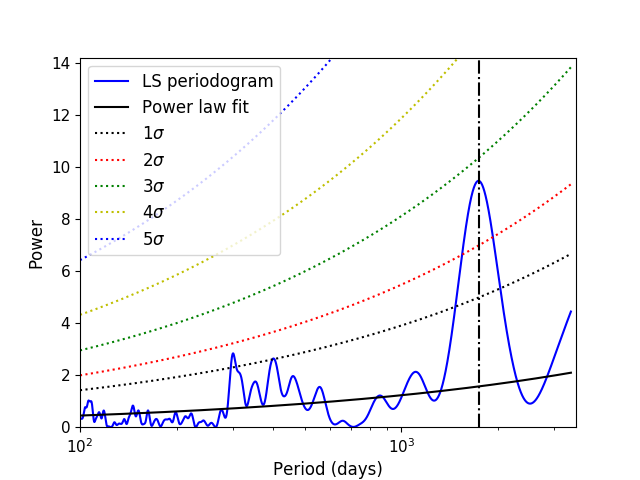}}
\subfigure{\includegraphics[width=0.84\columnwidth]{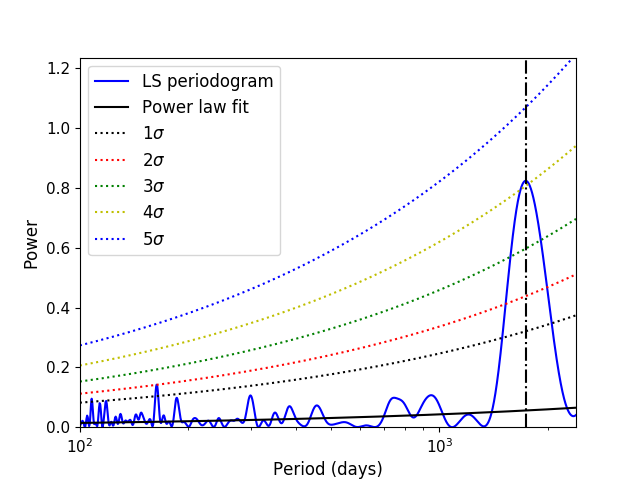}}
\subfigure{\includegraphics[width=0.83\textwidth]{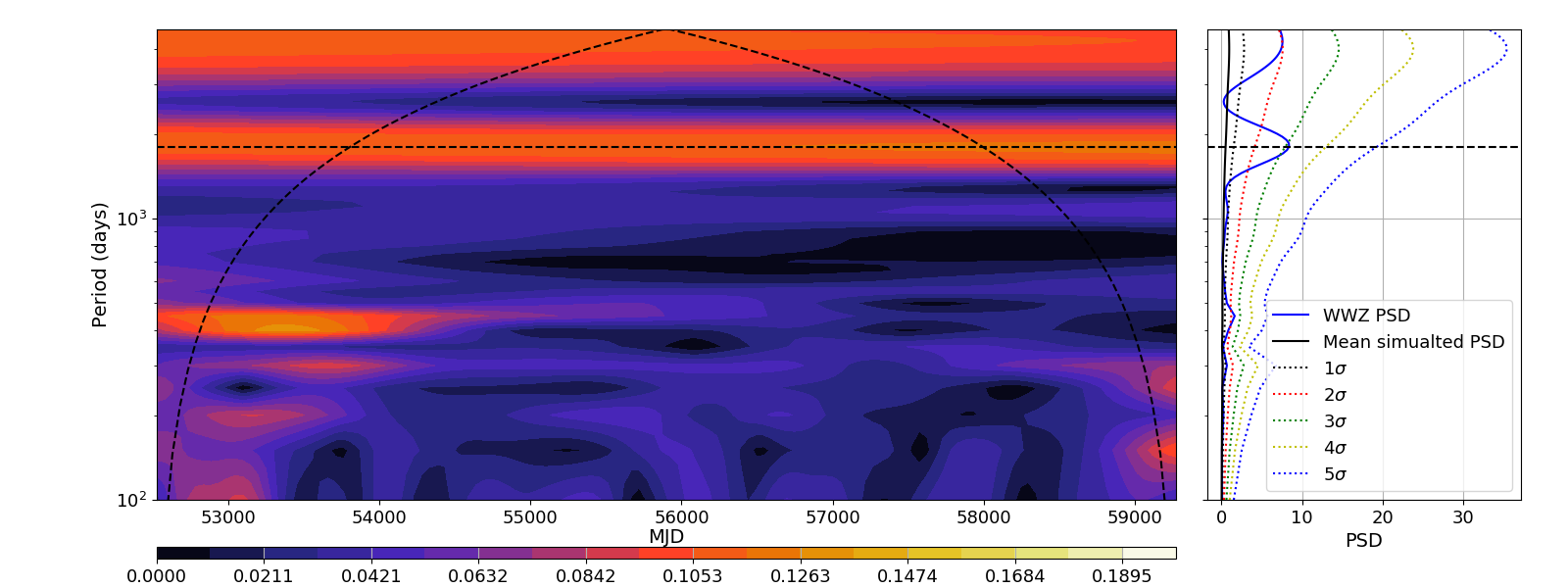}}
\subfigure{\includegraphics[width=0.83\textwidth]{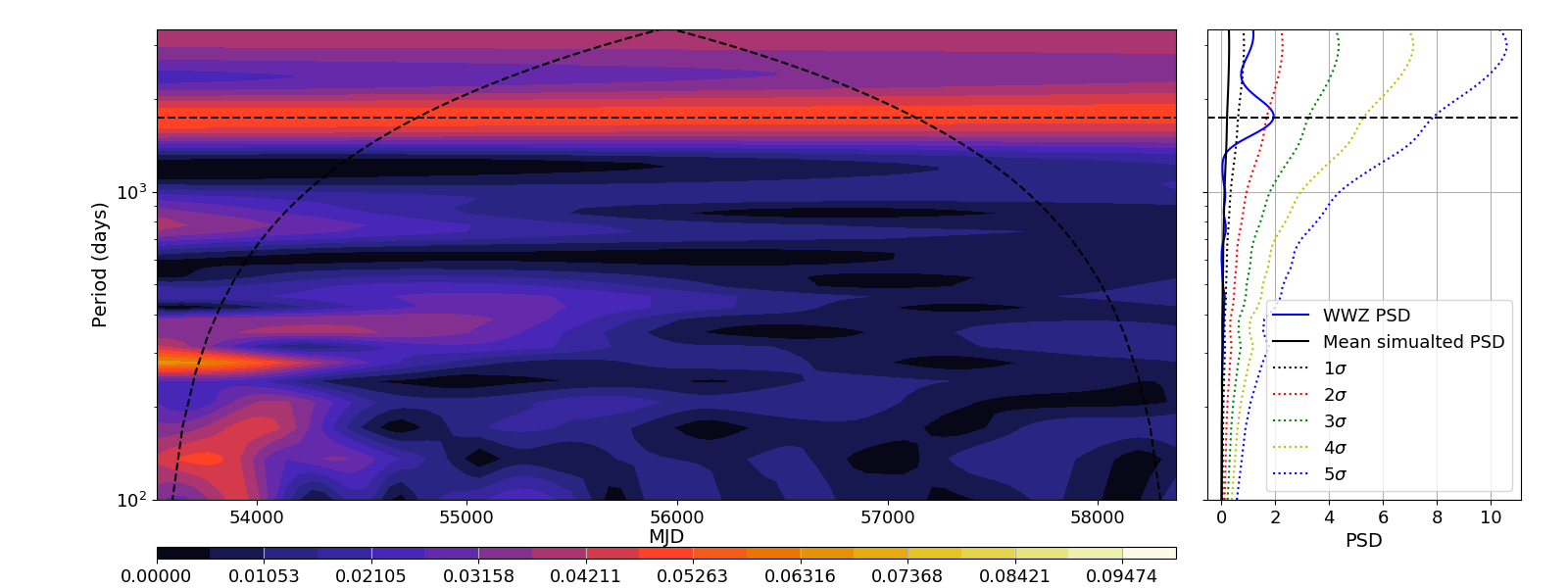}}
\caption{Periodicity analysis of the R- and V-band data of Mrk 501. \textit{First row}: ZDCF in R (left) and V (right) bands. \textit{Second row}: Lomb-Scargle periodogram in R (left) and V (right) bands. \textit{Third row}: WWZ diagram in the R band. \textit{Fourth row}: WWZ diagram in the V band. The left panels show the power spectrum as a function of period and time. The black dashed curves represent the COI. The right panels show the PSD. The coloured dotted lines represent the different significance levels. The black horizontal dashed line marks the peak of the PSD.}
\label{mrk501_results}
\end{figure*}

Moreover, the OVRO radio LC yields compatible results with those from the optical emission. The ZDCF, LS periodogram and WWZ derive a period of $4.7 \pm 0.5$~years (2.0$\sigma$-3.0$\sigma$), $5.7 \pm 1.0$~years (2.3$\sigma$) and $5.2 \pm 0.9$~years (2.1$\sigma$), respectively. The results for this band can be seen in Figure \ref{mrk501_results_radio}. The significance for the radio band is clearly lower than that derived for the optical bands due to the shorter time span. Therefore, we cannot strongly claim that this {QPO} appears in the radio emission of Mrk~501. However, the results reported here are promising and indicate that this period could be present also in this band. Finally, the HE $\gamma$-ray LC shows no hint of the same {quasi-periodic} modulation observed in the optical band.
For the case of Mrk~501, the detection of a possible {QPO} in the \textit{Fermi}-LAT data is however affected by shifted high-energy peak displayed by this blazar, located at VHEs, where the LAT is not as sensitive.


\subsection{BL Lacertae}
The first analysis step performed on BL~Lacertae reveals this blazar as a potential {quasi-periodic} emitter AGN. The R- and V-band LCs present a possible {QPO} of 19~years. Moreover, the polarization degree shows a preliminary positive result at a period of $\sim$1.6~years. The HE $\gamma$-ray emission of BL Lacertae presents a peak in the PDM and REDFIT methods at $\sim$3~years. Finally, no 15-GHz radio data are available for performing the {quasi-periodicity} analysis.

The analysis performed with the artificial LCs confirms the presence of the {quasi-periodicity} present in the polarized optical emission. The ZDCF has a very defined sinusoidal shape with a period of $1.6 \pm 0.1$~years. The LS periodogram shows a power peak at the same period. Finally, the WWZ shows a persistent {quasi-periodic} signal during the entire time span of the data, also at a period of $1.6 \pm 0.2$~years. After applying the trial factors considered here, the global significance of these methods was found to be $\sim$1.0$\sigma$-2.0$\sigma$ for the ZDCF, 2.6$\sigma$ for the periodogram, and 2.4$\sigma$ for the WWZ, respectively. These results are presented in Figure \ref{bllac_results_pol}.

\begin{figure*}
	\includegraphics[width=0.805\textwidth]{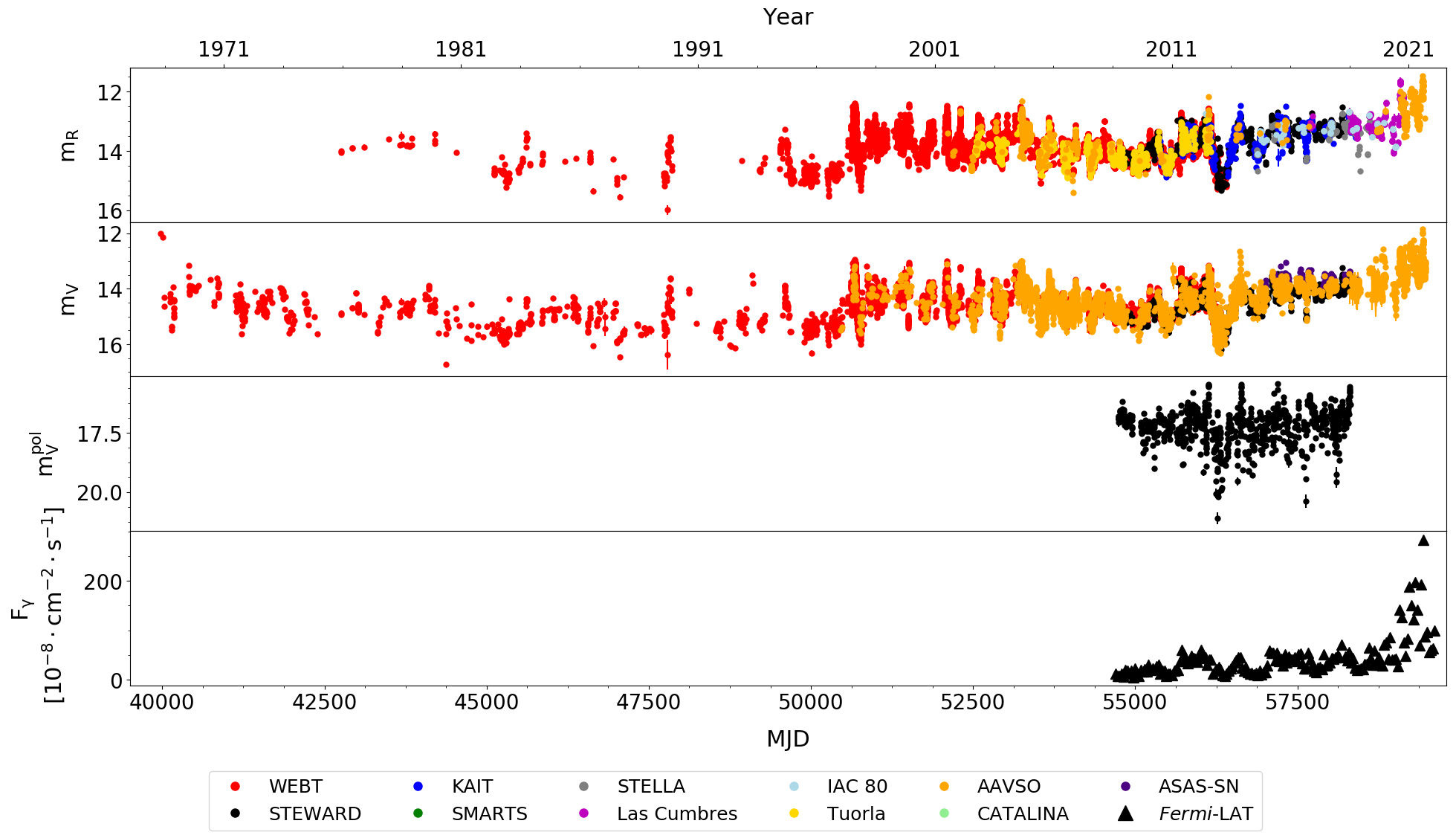}
    \caption{MWL LCs of BL Lacertae. Same description as Figure \ref{pks1222_lcs}.}
    \label{bllac_lcs}
\end{figure*}

\begin{figure*}
\centering
\subfigure{\includegraphics[width=0.85\columnwidth]{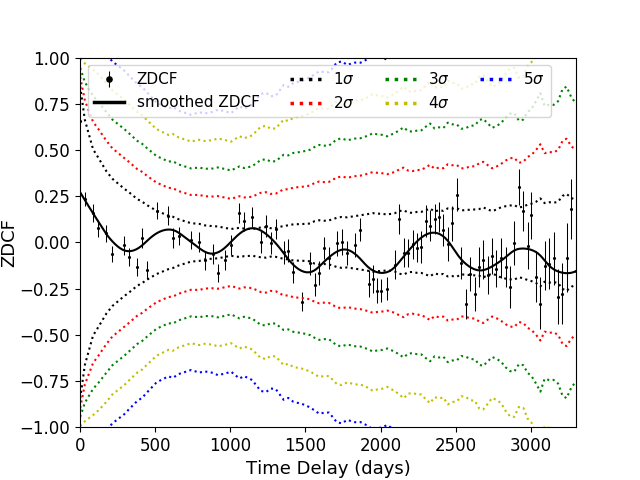}}
\subfigure{\includegraphics[width=0.85\columnwidth]{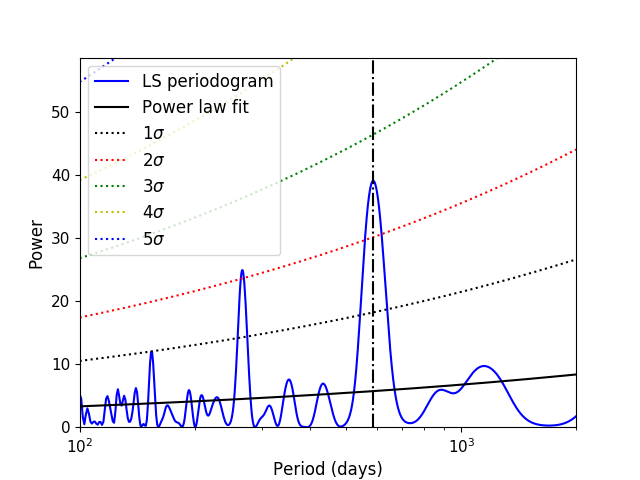}}
\subfigure{\includegraphics[width=0.85\textwidth]{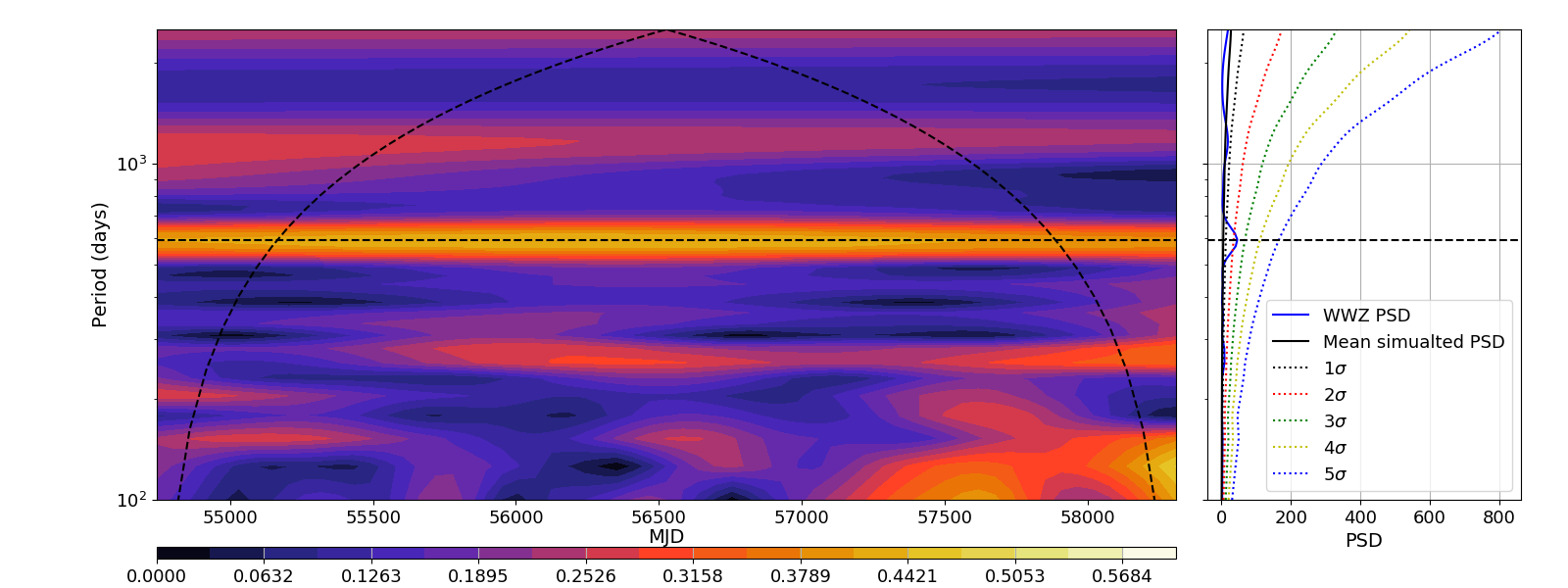}}
\caption{Periodicity analysis of the optical polarized data for BL~Lacertae. Same description as Figure \ref{ao0235_results}.} 
\label{bllac_results_pol}
\end{figure*}

Concerning the optical R and V bands, the {QPO} search shows a hint of a {quasi-periodic} modulation compatible with that found in the polarized light of BL~Lacertae in the WWZ, as well as the LS periodogram of the V-band LC. These methods report a period of $1.8 \pm 0.1$~years (0.9$\sigma$) and $1.8 \pm 0.2$~years (1.5$\sigma$) for the R-band LS periodogram and WWZ; and $1.8 \pm 0.1$~years (0.9$\sigma$) and $1.8 \pm 0.2$~years (1.8$\sigma$) for the V-band, in each method. These periodograms show a large variety of peaks at different frequencies due to different variability time scales. However, none of them shows a significance >1.5$\sigma$. Also, as revealed by the WWZ, these signals are not persistent in time and they most likely do not indicate real periodicities. The ZDCF on the other hand does not show a sinusoidal shape compatible with this period, as larger variability time scales dominate the correlation function, masking the impact of shorter variability time scales \citep[see e.g.][]{raiteri2021a,raiteri2021b}. In addition, the ZDCF shows a hint of a sinusoidal shape with a period $\sim$19~years, compatible with the one reported by the PDM and REDFIT methods (see Figure \ref{bllac_results}). The LS periodogram and WWZ also show a peak at a compatible value. However, after applying trial factors, we are not able to distinguish this potential signal from the underlying red noise, as derived from the low significance for this period (<1$\sigma$). Therefore, we cannot confirm the {quasi-periodic} nature of the longer signal. Despite this, BL Lac is still a promising target for presenting a {QPO} in its optical emission. The 19-year period has also been reported by \cite{fan1998} among other {quasi-periodic} signatures in the 100-year optical LC of this blazar. \cite{sandrinelli2017} also report a 680-day period for the R-band and $\gamma$-ray LCs, however not strongly confirmed. This latter period is compatible with the one reported for the polarization degree, as well as the shorter {QPO} found in the R and V bands. Finally, the HE $\gamma$-ray LC does not show any significant {quasi-periodic} modulation according to the analysis performed here.

\subsection{3C 454.3}
3C~454.3 is one of the most variable $\gamma$-ray blazars, with its emission and variability dominated by bright flares in all wavelengths, as observed from Figure \ref{3c454lc}. These flares clearly appear between MJD~52000-55500 and MJD~56500-58500, with a quiescent period in between both epochs. Moreover, the observations performed before MJD~52000 do not present any major flare or variability, as shown by the optical R- and V-band LCs. Due to the presence of these flares, some potential {QPOs} appear in the first analysis step. However, the ZDCF, LS periodogram and WWZ do not show any significant period in the broadband emission of this blazar after considering both the red noise and the trial factors (see Figure \ref{3c454_3_results}).

The optical polarization, radio and $\gamma$-ray LCs, with a much shorter time coverage than the optical data sets, yield a peak in the WWZ and LS periodogram at a period of $\sim$1100~days. This peak also appears in the WWZ of the optical data. However, as commented above, it is introduced by the series of flares observed in the LCs, as derived from the WWZ, where this peak is only present during the last third of the LC. The significance derived for this period in all cases is $\lesssim$1$\sigma$. Therefore, contrary to the results presented by \cite{li2006}, where two periods of 1.57~years and 6.15~years were reported, we conclude that the emission of this FSRQ is most likely dominated by stochastic flares rather than {quasi-periodic} enhanced flux states. 

\subsection{1ES 2344+514}
\begin{figure*}
	\includegraphics[width=0.805\textwidth]{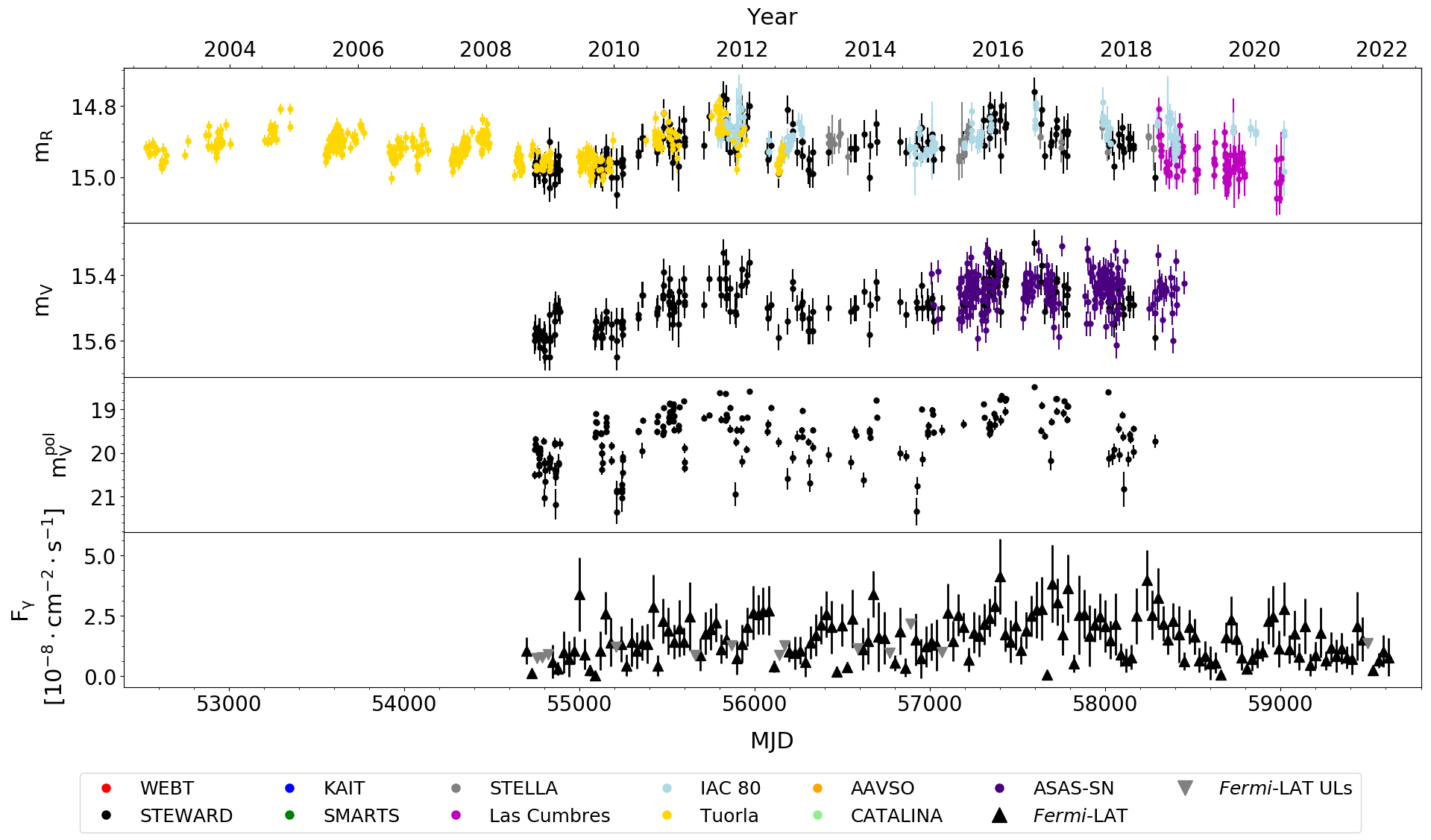}
    \caption{MWL LCs of 1ES 2344+514. Same description as Figure \ref{pks1222_lcs}.}
    \label{1es2344_lcs}
\end{figure*}

\begin{figure*}
\centering
\subfigure{\includegraphics[width=0.85\columnwidth]{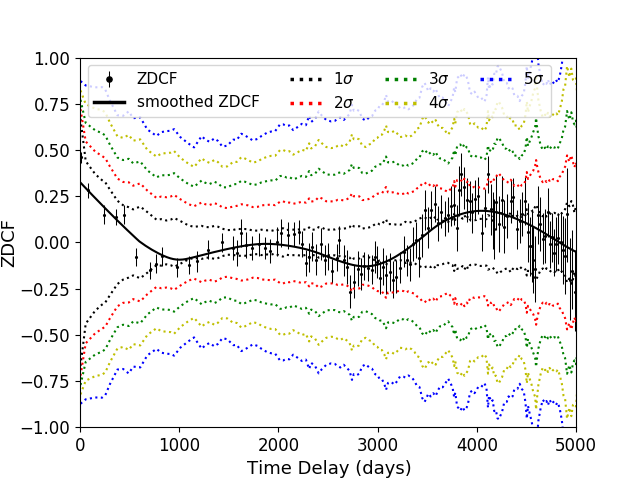}}
\subfigure{\includegraphics[width=0.85\columnwidth]{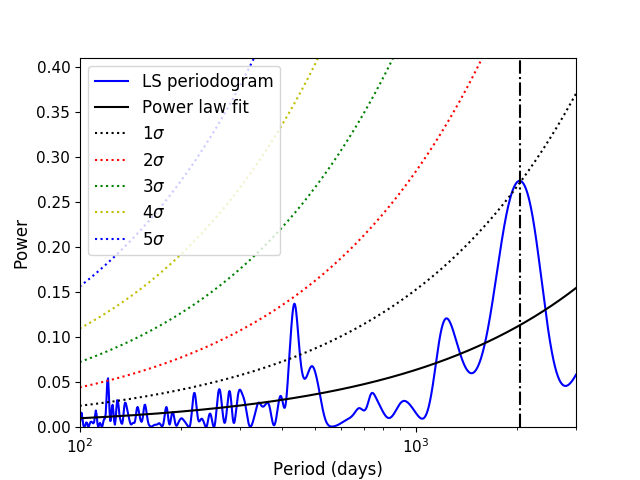}}
\subfigure{\includegraphics[width=0.85\textwidth]{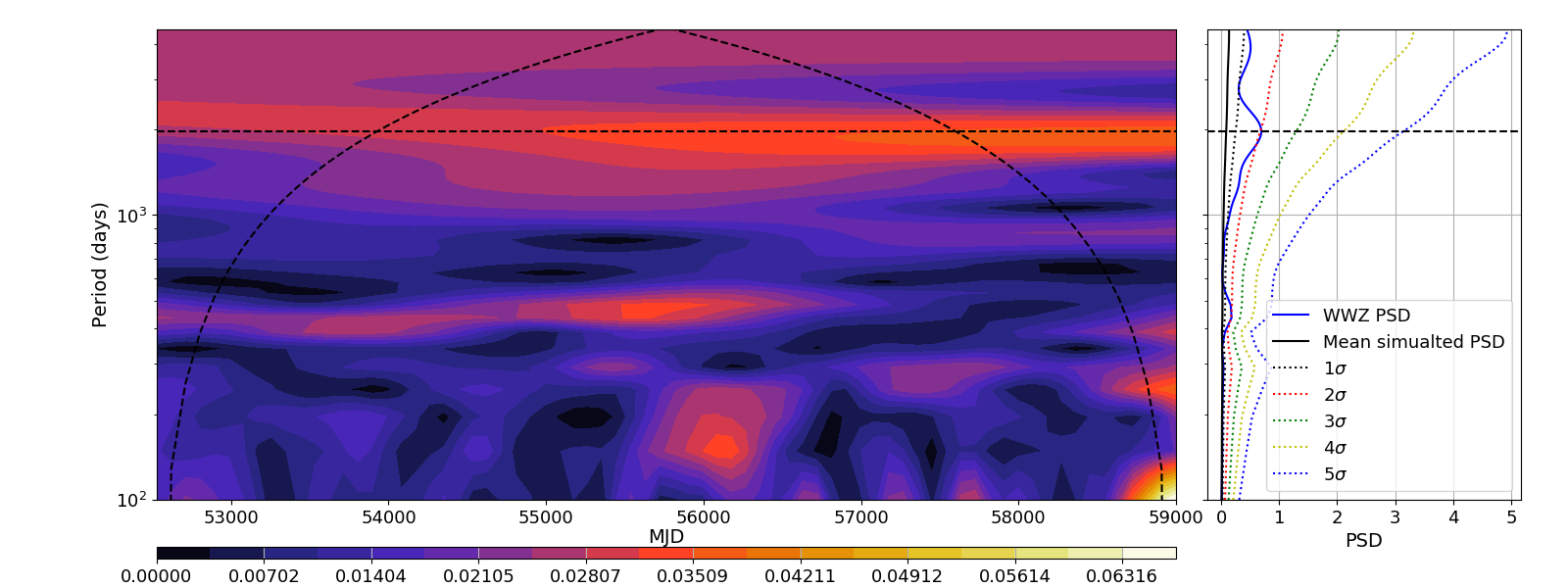}}
\caption{Periodicity analysis of the R-band data of 1ES 2344+514. Same description as Figure \ref{ao0235_results}.} 
\label{1es2344_results}
\end{figure*}

The REDFIT and PDM analyses show a possible {QPO} of $\sim$5~years in both the R and V optical bands, as well as the optical polarized emission of 1ES~2344+514. No hint or presence of {quasi-periodicity} is observed in the first analysis step in the $\gamma$-ray emission. Moreover, there are no radio data available for this object to test the {quasi-periodicity} in this band (see Figure \ref{1es2344_lcs}). The optical V-band and optical polarized LCs extend for roughly 10~years, approximately twice the value of the possible period. Therefore, extracting a reliable conclusion with such short time coverage is not possible in these bands. However, we have performed the detailed analysis in the optical R-band data. We observe a possible {QPO} with all three methods with a period of $\sim$5.6~years. The results are presented in Figure \ref{1es2344_results}.
This period is consistent with that from the preliminary methods. Nevertheless, the short time span of the data set w.r.t. the inferred period leads to a rather low significance of this result. Additionally, the LS periodogram and WWZ methods also show a period at $\sim$1.2~years, which could reveal an underlying faster period. This period reaches a significance of $\sim$2.0$\sigma$ in each method. However, it is masked in the ZDCF by the larger oscillation.

\begin{table*}
\centering
\caption{Results of the {quasi-periodicity} analysis for the sources with a significant positive detection above 2$\sigma$. The period measured in each band is reported in the corresponding column with its statistical significance. All the reported periods are expressed in years. All the confidence levels are already reporting the global significance, corrected by the trial factors.}
\label{periodicity_results}
\resizebox{\textwidth}{!}{%
\begin{tabular}{cccccccccccc}
\hline

\multirow{2}{*}{Source} & \multirow{2}{*}{Method} & \multicolumn{2}{c}{Optical R band} & \multicolumn{2}{c}{Optical V band} & \multicolumn{2}{c}{Optical polarization} & \multicolumn{2}{c}{HE $\gamma$-ray band} & \multicolumn{2}{c}{15-GHz radio band} \\ \cline{3-12} 
 &  &     Period      &      Sign.    & Period &     Sign.     &     Period      &    Sign.      &        Period   &   Sign.       &     Period      &      Sign.    \\ \hline
 
\multirow{3}{*}{AO~0235+164} & ZDCF & $8.2 \pm 0.3$ & 2.0$\sigma$ & -- & -- & -- & -- & -- & -- & -- & -- \\ 
 & LS & $8.3 \pm 0.6$ & 3.8$\sigma$ & -- & -- & -- & -- & -- & -- & -- & -- \\
 & WWZ & $8.7 \pm 1.5$ & 2.9$\sigma$ & -- & -- & -- & -- & -- & -- & -- & -- \\ \hline
 
\multirow{4}{*}{PKS 1222+216}  & \multirow{2}{*}{ZDCF} & $1.2 \pm 0.1$ & -- $^{a}$ & $1.2 \pm 0.1$ & -- $^{a}$ & $1.1 \pm 0.1$  & -- $^{a}$  & \multirow{2}{*}{--} & \multirow{2}{*}{--} & \multirow{2}{*}{--} & \multirow{2}{*}{--} \\
     &  & -- & -- & $4.7 \pm 0.3$ & 1.0$\sigma$-2.0$\sigma$ & $5.5 \pm 0.8$ & 1.0$\sigma$-2.0$\sigma$ &  &  &  &  \\
 & LS & $1.3 \pm 0.2$ & 1.3$\sigma$ & $1.3 \pm 0.1$ & 0.8$\sigma$ & $1.2 \pm 0.1$ & 3.3$\sigma$ & -- & -- & -- & -- \\ 
 & WWZ & $1.3 \pm 0.2$ & 1.4$\sigma$ & $1.3 \pm 0.2$ & 2.0$\sigma$ & $1.3 \pm 0.1$ & 3.0$\sigma$ & -- & -- & -- & -- \\ \hline
 
\multirow{3}{*}{Mrk~501} & ZDCF & $5.3 \pm 0.8$ & 2.0$\sigma$-3.1$\sigma$ & $4.8 \pm 0.6$ & 1.5$\sigma$-3.0$\sigma$ & -- & -- & -- & -- & $4.7 \pm 0.5$ & 2.0$\sigma$-3.0$\sigma$ \\ 
 & LS & $4.7 \pm 0.5$ & 2.8$\sigma$ & $4.8 \pm 0.4$ & 4.1$\sigma$ & -- & -- & -- & -- & $5.7 \pm 1.0$ & 2.3$\sigma$ \\  
 & WWZ & $5.0 \pm 0.8$ & 3.1$\sigma$ & $4.8 \pm 0.8$ & $2.2\sigma$ & -- & -- & -- & -- & $5.2 \pm 0.9$ & 2.1$\sigma$ \\ \hline
 
\multirow{3}{*}{BL Lacertae} & ZDCF & $20.5 \pm 1.2$ & 1.0$\sigma$-2.0$\sigma$ & $14.8 \pm 1.0$ & 1.0$\sigma$ & $1.6 \pm 0.1$ & 1.0$\sigma$-2.0$\sigma$ & -- & -- & -- & --\\  
 & LS & $1.8 \pm 0.1$ & 0.9$\sigma$ & $1.8 \pm 0.1$ & 0.9$\sigma$ & $1.6 \pm 0.1$ & 2.6$\sigma$ & -- & -- & -- & -- \\ 
 & WWZ & $1.8 \pm 0.2$ & 1.5$\sigma$ & $1.8 \pm 0.2$ & 1.8$\sigma$ & $1.6 \pm 0.2$ & 2.4$\sigma$ & -- & -- & -- & -- \\  \hline
 
  \multirow{3}{*}{1ES 2344+514} & ZDCF & $5.6 \pm 0.4$ & 1.0$\sigma$-2.0$\sigma$ & -- & -- & -- & -- & -- & -- & -- & --\\  
 & LS & $5.6 \pm 1.1$ & 1.0$\sigma$ & -- & -- & -- & -- & -- & -- & -- & -- \\ 
 & WWZ & $5.4 \pm 1.2$ & 2.0$\sigma$ & -- & -- & -- & -- & -- & -- & -- & -- \\  \hline
\end{tabular}%
}
\begin{flushleft}
\footnotesize{$^{a}$The dominance of the slow variability over the reported periods does not allow to extract their significance with the ZDCF.}
\end{flushleft}
\end{table*}

\section{Interpretation}\label{sec5}
Periodic and/or quasi-periodic oscillations in blazars have received numerous interpretations in the past few years. Depending on the characteristics of this periodic variability (e.g. single or double outburst, colour evolution or polarization variability), the different interpretations and models can provide useful information concerning the nature of the jet and the source. Two of the most employed ones consist in the presence of a supermassive binary black hole (SBBH) system in the heart of the AGN \citep[see for instance][]{sillanpaa1988, sillanpaa1996, oneill2022}, or the development of an helical jet or an helical motion inside the jet \citep[e.g.][]{rieger2004}.

In the framework of a SBBH system, a recurrent modulation of the emission is expected, caused by the influence of the secondary black hole on the primary. Several scenarios have been proposed within this interpretation, depending on the effect caused by the secondary black hole:
\begin{enumerate}
    \item The original SBBH model uses tidal perturbations in the primary accretion disk introduced by the smaller black hole as the cause of the observed periodicity. This produces a periodic gas transfer towards the primary black hole, leading to the observed periodicity. This model has been originally applied to OJ~287 and its $\sim$12-year period \citep{sillanpaa1988}. However, it cannot reproduce the double-peak structure observed in this object.
    
    \item The {(quasi-)}periodicity occurs due to perturbations in the accretion rate caused by the secondary black hole piercing the accretion disc of the primary. During its orbital motion around the primary, each time the secondary crosses the primary accretion disc, it induces changes in the accretion rate that lead to recurrent major flares \citep{lehto1996}. This model predicts a double-peak structure during the outburst phase, as observed for the most studied periodic blazar OJ~287 \citep[see][]{sillanpaa1996}.
    
    \item Each black hole hosts a relativistic jet that contributes to the overall emission \citep{villata1998,huang2021}. The {(quasi-)}periodic variability is then introduced by the orientations of both jets. The double peak structure observed in some cases is explained as the consecutive contributions of both jets.
    
    \item The {(quasi-)}periodic emission is caused by microlensing effect introduced by a massive object or the secondary black hole moving between the primary and the observer \citep[see][]{sillanpaa1996}. Under this interpretation, recurrent and achromatic outbursts are expected. Additionally, for the latter case, very symmetric flares are also predicted.
    
    \item A bent or twisted helical jet due to the orbital movement of the companion around the primary black hole \citep{villata1999}. A {(quasi-)}periodic emission is expected in this scenario, introduced by the different intensity of the relativistic beaming due to the changing viewing angle of the emitting region along the helical path \citep{ostorero2004,sobacchi2017}. 
    
    \item The motion of the secondary black hole leads to changes in the velocity of the material in the jet relative to the observer \citep{oneill2022}. Hence, this induces changes in the Doppler factor, causing the sinusoidal variability.
    
\end{enumerate}
A detailed discussion on the first four scenarios can be found in \cite{sillanpaa1996} and references therein. The latter two are discussed by \cite{villata1999} and \cite{oneill2022}, respectively. 

Each of these SBBH models would produce characteristic signatures in the emission and the {quasi-periodicity} observed in the data. For instance, similarly to OJ~287 with its $\sim$12-year period \citep[][]{sillanpaa1988,sillanpaa1996}, AO~0235+164 was interpreted in the past as a SBBH system \citep[e.g.][]{raiteri2001,ostorero2004,roy2022}. A visual inspection of the inner panel of Figure \ref{ao0235_lcs} shows a highly symmetric double peak structure of the recurrent flares along the R-band LC. In this framework, the model applied to OJ~287 based on the perturbations introduced in the accretion disc and transferred to the jet, could reproduce the double peaks and the MWL {QPO}. However, given the thermal origin of the flares under this interpretation, no periodic emission is expected for the polarization degree, contrary to what it seems to be observed here (nevertheless with a very short time coverage), as no mechanism involving the magnetic field would be responsible for the observed {QPO}. Alternatively, a binary black hole model based on helical jets such as the one proposed by \cite{villata1998,ostorero2004}, where the magnetic field also follows a helical structure \citep{raiteri2013}, can explain the recurrent emission in the broadband emission, as well as the polarized flux. The double peak would be caused by the precession of the jet of each black hole \citep[see][]{villata1998}. 

\cite{roy2022} propose a second plausible interpretation based on SBBH systems, where the recurrent bright flares would be produced by gravitational lensing effect. In this scenario, the lensing effect could be caused by the secondary black hole moving between the primary and the line of sight during its orbit around the bigger black hole \citep{sillanpaa1996}. This model would explain the presence of the symmetric double peaks appearing in the broadband emission of this source. Under this interpretation, these flares are expected to be mainly achromatic, as no physical change is behind the brightening. This achromatic brightening was reported by \cite{otero-santos2022}. This scenario can be compatible with the correlation observed in the polarized emission. The correlation between the different bands without time delay indicates that the MWL emission is being produced in the same region of the jet \citep{max-moerbeck2014}. If the emission of this region is being amplified via lensing effect, and the region is polarized, then the polarization degree would also be higher as the polarized emission would be boosted by the lensing effect. Therefore, the same behaviour is expected in the optical polarization.

The binary black hole scenario could also be a plausible explanation for other sources. \cite{dey2018} observed for OJ~287 the presence of a faster and slower periodicity. Here we cannot confirm the 1.6-year and $\sim$19-year periodicities for the total emission of BL~Lacertae. However, if these periods can be confirmed, a binary black hole model could provide a potential solution for understanding this object. In the same way, this could be applied to Mrk~501. \cite{bhatta2021} reported for this source a shorter period of 333~days. This period is not clearly observed here, however this could be due to the dominance of the slower 5-year time scale variation \citep{raiteri2021a,raiteri2021b}. Therefore, if both periods can be confirmed, this could be interpreted as a binary system. Finally, the two faint hints observed in 1ES~2344+514 can also favour this scenario if confirmed in the future.

On the other hand, another common interpretation is based on geometrical effects \citep{rieger2004} such as the presence of {a twisted} or helical jet and/or magnetic field \citep[see the model proposed by][]{raiteri2013}, that could explain the presence of {quasi-periodic} variability in the total and polarized emission, respectively \citep[see e.g.][]{raiteri2013}. These helical or twisted paths along the jet produce changes in the Doppler factor and therefore, in the relativistic Doppler boosting. The relativistic Doppler factor $\delta$ can be defined as
\begin{equation}
\delta=\frac{1}{\Gamma[1-\beta\cos(\theta)]},
\label{doppler_factor}
\end{equation}
with $\Gamma$ being the Lorentz factor, $\beta$ the velocity of the jet in units of the light speed and $\theta$ the viewing angle. {Quasi-}periodic changes in the viewing angle lead to potential {QPOs} in the emission of the sources. Moreover, these models are also able to reproduce and interpret {quasi-}periodic changes in the polarization degree of blazars, in comparison with models based on disc instabilities, where there is no reason to expect any correlation between the total and polarized emission. \cite{raiteri2013} proposed a model aiming to explain and reproduce the variability observed in the polarized flux of BL Lacertae. This model assumes the existence of a helical magnetic field where the polarization degree is modulated by the viewing angle, similarly to the helical jet framework, as
\begin{equation}
p=p_{max}\sin^{2}(\theta '),
\label{P_helical_jet}
\end{equation}
where $\theta '$ is the viewing angle in the rest frame of the jet, related to the observed angle as
\begin{equation}
\sin(\theta ')=\frac{\sin(\theta)}{\Gamma[1-\beta\cos(\theta)]}.
\label{theta_prima}
\end{equation}
The changes in the viewing angle can be derived assuming that the {long-term} variability comes from Doppler factor changes {of geometrical origin, {while short-term variability would be most likely ascribed to intrinsic energetic processes.} The Doppler factor can be derived from the flux densities according to}
\begin{equation}
\delta=\delta_{max}(F/F_{max})^{1/(n+\alpha)}, 
\label{delta_deltamax}
\end{equation}
where {$\alpha$ is the intrinsic spectral index and} $\delta_{max}$ is defined through the minimum viewing angle using Equation (\ref{doppler_factor}). {For a continuous jet, $n=2$ as reported by \cite{urry1995}. Moreover, following the modeling from \cite{raiteri2013}, we adopt $\alpha=1$.} This model can lead to {quasi-}periodic changes of the polarization degree with changes of the viewing angle \citep[see Figure 17 from][]{raiteri2013}. {As a first approximation, the jet under this model can be interpreted as a rotating helix. Nevertheless, in a more general scenario it is most likely a chaotic twisting jet composed of twisting filaments rather than a rigid structure, causing perturbations to the perfect periodic behaviour.}

Another alternative of applying this model involves assuming an intrinsic periodic change of $\theta$ considering the measured period $P_{obs}$. In this scenario, the angle between the jet and the direction of the emitting region (pitch angle) $\phi$ and the angle between the jet and the line of sight $\psi$ can be estimated as
\begin{equation}
\cos(\theta)=\cos(\phi)\cos(\psi)+\sin(\phi)\sin(\psi)\cos(2\pi t/P_{obs}).
\label{cos_theta}
\end{equation}
Thus, a model based on helical jets can lead to {quasi-}periodic changes in both the total and polarized emission of blazars. Therefore, the presence of a helical jet developing a helical magnetic field could potentially explain the {long-term} periodicity in both the MWL and polarized emissions of the target.

\begin{figure}
	\includegraphics[width=\columnwidth]{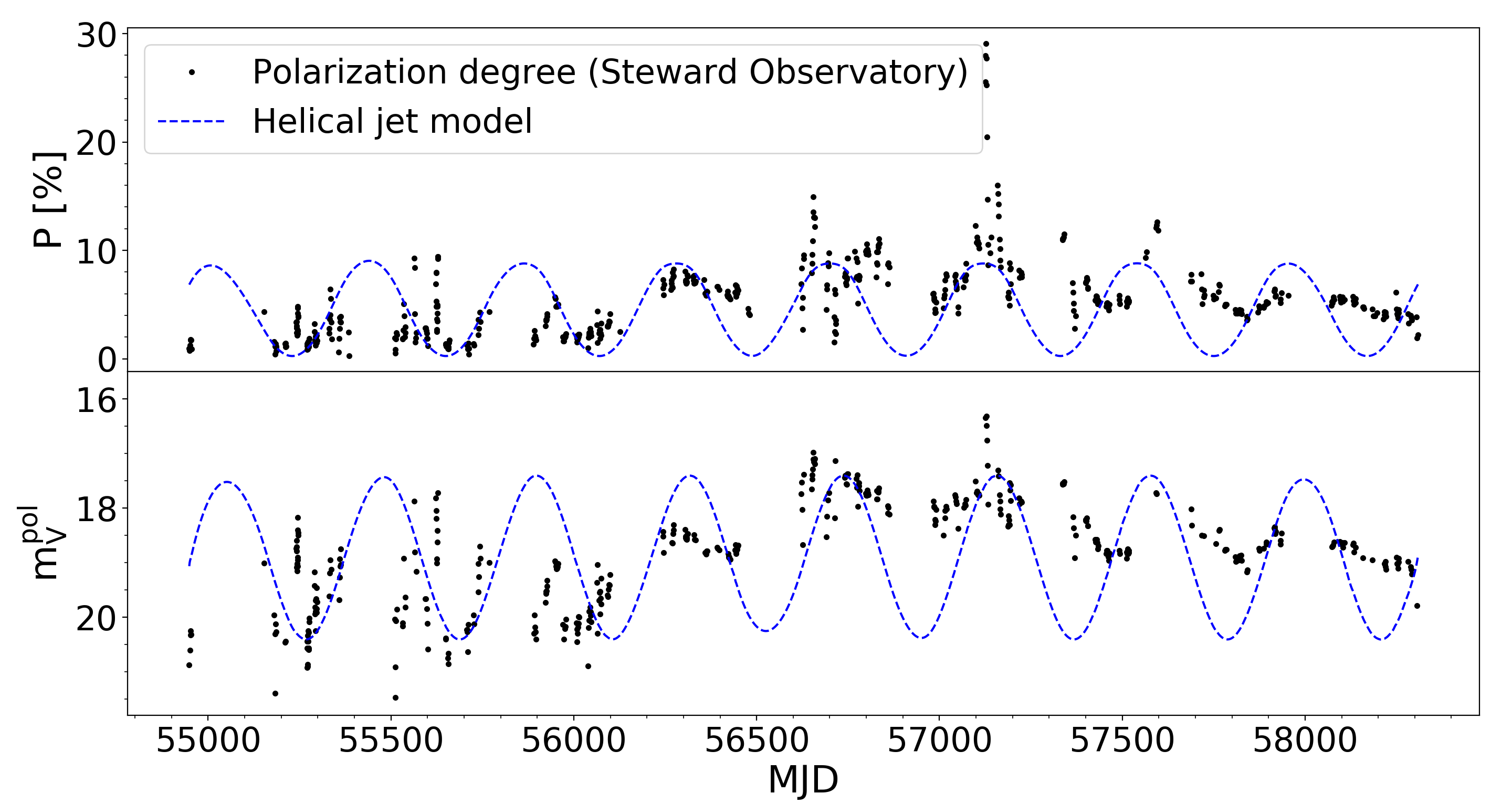}
    \caption{Helical jet model of PKS~1222+216. \textit{Top:} Helical jet model for the polarization degree. \textit{Bottom:} Helical jet model for the polarization degree transformed to polarized magnitude.}
    \label{pks1222_helical_jet}
\end{figure}

The helical jet scenario can provide a reasonable explanation for the observed long-term behaviour of PKS~1222+215, involving a helical jet with a magnetic field following the same helical structure. As a first approach, we assume that the long-term flux variability comes from changes in the Doppler factor using Equation (\ref{delta_deltamax}). We have tested this model for different values of $\Gamma$, $\theta_{min}$ and $\delta_{max}$. However, we are not able to reproduce the observed polarization changes and the range of its variability. Therefore, the helical jet model assuming that the long-term polarization variability is caused by changes in the Doppler factor is not compatible with our observations. Alternatively, we have tried a helical jet model under the assumption that the variations of the viewing angle are intrinsically {quasi-}periodic (due to, for instance, helical movements of the emitting blob or region), following Equation (\ref{cos_theta}). The polarization degree then changes with the viewing angle following Equation (\ref{P_helical_jet}), while the total flux is modulated by {quasi-}periodic changes in the Doppler factor $\delta$, responsible of the relativistic boosting. We have used the parameters derived from radio observations $\Gamma=13.9$ and $\phi=1.6^{\circ}$, presented by \cite{jorstad2017}. Iterating over different values of the angle $\psi$, we find a reasonably good agreement between the data and the model for $\psi=6^{\circ}$, as presented in Figure \ref{pks1222_helical_jet}. Under this assumption, the viewing angle of the emitting blob or region oscillates between $\sim$4$^{\circ}$ and $\sim$7.5$^{\circ}$, compatible with the mean value of $\langle\theta\rangle=5.6^{\circ}$ measured by \cite{jorstad2017}. The {quasi-}periodic modulation expected from the theoretical model can also be observed in the real measurements of the optical magnitude and the optical polarization degree. We note however that the intensity of each maximum and minimum is not always well reproduced. This is expected as this approach is only able to reproduce the sinusoidal variability observed. Shorter variability time scales impacting the development of the LC are expected to be responsible for the different intensity of each flare and oscillation. {Another alternative involves considering a variable pitch angle with time to reproduce the amplitude of the major flares \citep[see for instance][]{raiteri2021b}.} Nevertheless, this scenario provides a plausible interpretation for the observed {quasi-}periodicity. We note that, as stated by \cite{raiteri2013}, {this helical or twisted jet could be caused by either the presence of a SBBH system or a jet precession due to e.g. a misalignment between the normal to the accretion disk and the supermassive black hole axis}. \cite{zhang2022} have also recently applied this model with success, finding similar parameter values.

We have also tested this model for Mrk~501. As for PKS~1222+216, we are not able to reproduce the amplitude of the variability assuming that changes in the Doppler factor lead to the observed long-term variability. Also, using the typical $\Gamma$ values reported in the literature for this source \citep[e.g. $\Gamma=5$ by][]{giovannini2001}, the model predicts an anticorrelated behaviour of the polarization degree with the flux, not observed in the data available here. Therefore, we also applied the approach of assuming an intrinsic {quasi-}periodic modulation of the viewing angle following Equation (\ref{cos_theta}). Under this scenario, we obtained an approximated reproduction of the {QPO} using $\Gamma=5$, $\phi=5^{\circ}$ and $\psi=15^{\circ}$, following the expected oscillation of $\theta$ between $\sim$10-15$^{\circ}$ from \cite{giovannini2001}. The model predicts a similar angle variation of $\sim$10-20$^{\circ}$. However, this interpretation also predicts a {quasi-}periodic behaviour of the polarization as indicated in Equation (\ref{P_helical_jet}), not observed here in the polarized magnitude.

\begin{figure}
    \centering
	\includegraphics[width=0.82\columnwidth]{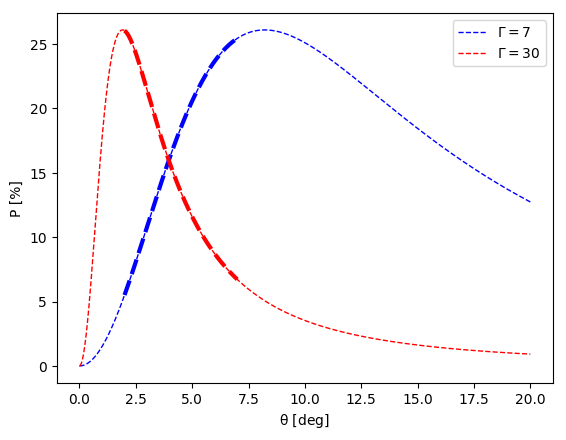}
    \caption{Helical magnetic field model of BL~Lacertae for $\Gamma=7$ (blue) and $\Gamma=30$ (red). The variability range of $\theta$ predicted by the model is highlighted with a thicker line.}
    \label{bllac_helical_jet}
\end{figure}

Finally, the {quasi-}periodicity observed in the polarized emission of BL~Lacertae can also be evaluated within this scenario. \cite{raiteri2013} used the helical jet model to reproduce the long-term variability and anticorrelated behaviour observed between the flux and the polarization degree. The data sample used by \cite{raiteri2013} for measuring this anticorrelation corresponds to roughly the first half of our data set (MJD~54550 to MJD~56200, approximately). While the anticorrelation reported by these authors can be observed for the period covered by both data sets, we observe some correlated behaviour for the latest observations, not included in the former study. We adopt $\Gamma=7$ from \citep{jorstad2005} and $\theta_{min}=2^{\circ}$ reported by \citep{larionov2010}, following the approach from \cite{raiteri2013}, and assuming that {the long-term variability} is introduced by Doppler factor changes as defined by Equation (\ref{P_helical_jet}). This model can roughly predict the behaviour of the polarization degree for the anticorrelated part of the LC. However, it fails to explain the polarization changes for the most recent observations, where the correlation is observed. This change in the {long-term} tendency of the polarization degree and the flux can be understood as an increase of the Lorentz factor $\Gamma$. As observed from Figure 17 in \cite{raiteri2013} and Figure \ref{bllac_helical_jet}, a change from a lower to a higher $\Gamma$ (e.g. $\Gamma=30$) for the same oscillation of $\theta$ could lead to a change from the anticorrelated to the correlated part of the curve. 

{We note that this model is performed by applying a 60-day binning and a cubic spline interpolation of the LC, following the prescriptions from \cite{raiteri2013}. Owing to the rather short 1.6-year period observed for this source, the binning and interpolation can lead to a smoothing of features in the LC. Therefore, the {quasi-}periodicity may appear somewhat diluted in the derived model. However, the model is able to roughly reproduce the long-term variability of the source both in total and polarized flux for the two Lorentz factors considered. Moreover, when applying the LS periodogram to the model of each half of the LC, a hint of a 1.6-year period compatible with the measured {quasi-}periodic signal appears. This hint shows a low significance (<2$\sigma$) due to the shorter time coverage of each half of the LC and the effect of the smoothing. However, we find consistent results for the modeling and the real observations. Hence, this scenario can provide a reasonable explanation to the observed behaviour.}

We also stress that this blazar presents a very variable emission in very different time scales, as observed in the MWL LCs (see Figure~\ref{bllac_lcs}), and as reported in the past in previous variability analyses \citep[see for instance][]{gaur2015a,gaur2015b}. The variability in the shortest and longest time scales could be masking the impact of this 1.6-year {quasi-}periodicity and preventing us to detect the underlying faster variability in the R- and V-band LCs with a higher significance. {In addition, this model only intends to explain the origin of the long-term variations of the source, while shorter time scales could be associated with intrinsic energetic processes.}

\section{Conclusions}\label{sec6}
We have performed an extensive {quasi-}periodicity search in the MWL emission of several $\gamma$-ray emitting blazars monitored by the Steward Observatory. We performed the analysis in the blazar sample with enough temporal extension in their optical total and polarized emission. The optical data sets used here have an incredibly long time coverage, with $\sim$15-20~years of data and reaching up to $\sim$40-60~years for some sources. Moreover, we have collected MWL data with the \textit{Fermi}-LAT $\gamma$-ray LCs from the public LCR, as well as the 15-GHz radio data from OVRO. Among the sources included in this work, we have detected several promising candidates for presenting {quasi-}periodic variability in their emission in one or more bands.
\begin{itemize}
    \item AO 0235+164 presents a possible {quasi-}periodic modulation in its optical R-band emission of $\sim$8.2~years with a statistical significance $\sim$3$\sigma$ after trial corrections. However, the temporal coverage of the V-band, optical polarization, $\gamma$-ray and radio data gathered does not allow to reliably test the existence of this period. Nevertheless, the correlated appearance of the bright flares could indicate that this period is also present in the latter wave bands.

    \item PKS 1222+216 presents solid evidence ($\sim$3$\sigma$) of a {quasi-}periodic modulation in its polarized emission, with a period of $\sim$1.2~years. In addition, the R and V bands show a hint of the same period in the total optical emission, with a lower significance. However, the presence of two intense flares may be masking this {QPO}, as observed in the results for the $\gamma$-ray LC of this blazar.
    
    \item A promising period of $\sim$5~years was observed in the optical and radio emission of Mrk~501, with significances $\sim$3$\sigma$ and even reaching 4$\sigma$ with some of the methods. No significant presence of this period appears in the polarized light or the $\gamma$-ray emission. However, we note that for the latter, the low temporal coverage of the \textit{Fermi}-LAT data (spanning for roughly 2 cycles) and the low sensitivity at the energies around the highest $\gamma$-ray emission of this source could be severely affecting the analysis.
    
    \item BL Lacertae shows a {quasi-}periodic signature in its optical polarized emission, with a period of $\sim$1.6~years and a confidence level of $\sim$2.5$\sigma$. Moreover, the optical R- and V-band total emission also shows a hint of a compatible period ($\sim$1.8~years), also observed by \cite{sandrinelli2017}. However, these bands show a lower significance than the polarized data set. In addition, these optical bands display a strong peak compatible with a period of $\sim$19~years, reported in the past by \cite{fan1998}. The long nature of this possible {quasi-}periodicity, approximately half of the span of the R-band data set, and one third of the V-band data, does not allow to distinguish it from the inherent red noise to time series of blazars.
    
    \item The R-band emission of 1ES 2344+514 presents a weak ($\sim$2$\sigma$) {QPO} hint with a value of $\sim$5.5~years. Moreover, a possible faster period ($\sim$1.2~years) is also reflected in the R-band LS periodogram and WWZ, indicating a second possible {quasi-}periodic modulation. The optical V and polarized bands, as well as the HE $\gamma$-ray band, do not have the temporal coverage needed to reliably test this period.
\end{itemize}

We have explored several scenarios that could lead to the appearance of (quasi-)periodicities in the MWL emission of blazars. We have interpreted the results found here with models based on a SBBH system that can cause the recurrent emission through different mechanisms: instabilities transmitted from the accretion disc to the jet, helical jets, or gravitational lensing effect \citep{sillanpaa1996}. Moreover, scenarios based on geometrical effects like helical paths of instabilities propagating along the jet and helical magnetic fields or jets with a twisted structure have also been considered, successfully explaining the {quasi-}periodic emission observed when both the total and polarized flux show such behaviour \citep{raiteri2013}.

We have identified evidences of {quasi-}periodic variations in our blazar sample that can be well described under different scenarios, depending on the characteristics of their broadband emission. However, the long-term nature of some of the periods found here, the short time coverage of some data samples, and the sometimes low significance do not permit to robustly claim the existence of some of the results in the MWL emission (e.g. AO~0235+164 with a $\sim$8.2-year period, or PKS~1222+216 with a low statistical significance in its total optical emission). Extending the MWL data sets for these blazars will be crucial to further understand the origin and nature of this {quasi-}periodic variability.

\section*{Acknowledgements}
JOS thanks the support from grant FPI-SO from the Spanish Ministry of Economy and Competitiveness (MINECO) (research project SEV-2015-0548-17-3 and predoctoral contract BES-2017-082171). 
JOS, JAP and JBG acknowledge financial support from the Spanish Ministry of Science and Innovation (MICINN) through the Spanish State Research Agency, under Severo Ochoa Program 2020-2023 (CEX2019-000920-S). 
PP acknowledge funding under NASA contract 80NSSC20K1562. Data from the Steward Observatory spectropolarimetric monitoring project were used. 
This program is supported by Fermi Guest Investigator grants NNX08AW56G, NNX09AU10G, NNX12AO93G, and NNX15AU81G. 
This work is partly based on data taken and assembled by the WEBT collaboration and stored in the WEBT archive at the Osservatorio Astrofisico di Torino - INAF (\url{https://www.oato.inaf.it/blazars/webt/}).
This paper has made use of up-to-date SMARTS optical/near-infrared light curves that are available at \url{www.astro.yale.edu/smarts/glast/home.php}.
This research has made use of data from the robotic 0.76-m Katz- man Automatic Imaging telescope \citep{li2003} at Lick Observatory.
This work makes use of data from the Catalina Real-Time Transient Survey (CRTS) and the All-Sky Automated Survey for Supernovae (ASAS-SN).
This work makes use of observations from the Las Cumbres Observatory global telescope network.
Based (partly) on data obtained with the STELLA robotic telescopes in Tenerife, an AIP facility jointly operated by AIP and IAC.
This article is based on observations made with the IAC80 operated on the island of Tenerife by the Instituto de Astrofísica de Canarias in the Spanish Observatorio del Teide.
We acknowledge with thanks the variable star observations from the AAVSO International Database contributed by observers worldwide and used in this research.
This research has made use of data from the OVRO 40-m monitoring program \citep{richards2011} which is supported in part by NASA grants NNX08AW31G, NNX11A043G, and NNX14AQ89G and NSF grants AST-0808050 and AST-1109911.
We thank the anonymous referee for his/her comments and review of the manuscript.

\section*{Data Availability}

All the data used in this work are publicly available or available on request to the responsible of the corresponding observatory/facility. All the links to the data bases, online repositories and/or contact information are provided in the footnotes in Section \ref{sec2}.



\bibliographystyle{mnras}
\bibliography{bibliography} 




\section*{Supplementary material}

\noindent\textbf{Figure A1\labtext{A1}{oj248lc}}. MWL LCs of OJ 248 (same description as Figure \ref{pks1222_lcs}, available in the online version).

\noindent\textbf{Figure A2\labtext{A2}{oj287lc}}. MWL LCs of OJ 287 (same description as Figure \ref{pks1222_lcs}, available in the online version).

\noindent\textbf{Figure A3\labtext{A3}{mrk421lc}}. MWL LCs of Mrk 421 (same description as Figure \ref{pks1222_lcs}, available in the online version).

\noindent\textbf{Figure A4\labtext{A4}{wcomlc}}. MWL LCs of W Comae (same description as Figure \ref{pks1222_lcs}, available in the online version).

\noindent\textbf{Figure A5\labtext{A5}{3c273lc}}. MWL LCs of 3C 273 (same description as Figure \ref{pks1222_lcs}, available in the online version).

\noindent\textbf{Figure A6\labtext{A6}{3c279lc}}. MWL LCs of 3C 279 (same description as Figure \ref{pks1222_lcs}, available in the online version).

\noindent\textbf{Figure A7\labtext{A7}{pks1510lc}}. MWL LCs of PKS~1510-089 (same description as Figure \ref{pks1222_lcs}, available in the online version).

\noindent\textbf{Figure A8\labtext{A8}{1es1959lc}}. MWL LCs of 1ES~1959+650 (same description as Figure \ref{pks1222_lcs}, available in the online version).

\noindent\textbf{Figure A9\labtext{A9}{cta102lc}}. MWL LCs of CTA 102 (same description as Figure \ref{pks1222_lcs}, available in the online version).

\noindent\textbf{Figure A10\labtext{A10}{3c454lc}}. MWL LCs of 3C 454.3 (same description as Figure \ref{pks1222_lcs}, available in the online version).

\noindent\textbf{Figure A11\labtext{A11}{pks1222_results}}. Periodicity analysis of the R- and V-band data of PKS~1222+216 (same description as Figure \ref{mrk501_results}, available in the online version).

\noindent\textbf{Figure A12\labtext{A12}{3c273_results}}. Periodicity analysis of the V-band data of 3C~273 (same description as Figure \ref{ao0235_results}, available in the online version).

\noindent\textbf{Figure A13\labtext{A13}{3c279_results}}. Periodicity analysis of the V-band data of 3C~279 (same description as Figure \ref{ao0235_results}, available in the online version).

\noindent\textbf{Figure A14\labtext{A14}{pks1510_results}}. Periodicity analysis of the R-band data of PKS~1510-089 (same description as Figure \ref{ao0235_results}, available in the online version).

\noindent\textbf{Figure A15\labtext{A15}{mrk501_results_radio}}. Periodicity analysis of the optical 15-GHz radio data for Mrk~501 (same description as Figure \ref{ao0235_results}, available in the online version).

\noindent\textbf{Figure A16\labtext{A16}{bllac_results}}. Periodicity analysis of the R- and V-band data of BL~Lacertae (same description as Figure \ref{mrk501_results}, available in the online version).

\noindent\textbf{Figure A17\labtext{A17}{3c454_3_results}}. Periodicity analysis of the R-band data of 3C~454.3 (same description as Figure \ref{ao0235_results}, available in the online version).







\bsp	
\label{lastpage}
\end{document}